\documentclass{article}
\usepackage[utf8]{inputenc}
\usepackage[a4paper, total={6in, 8in}, margin = 3cm]{geometry}
\usepackage[english]{babel}
\usepackage{authblk}
\usepackage{enumitem}
\usepackage{mathtools}
\usepackage{amssymb}
\usepackage{amsmath}
\usepackage{setspace}
\usepackage{natbib}
\usepackage{booktabs}
\usepackage{multirow}
\usepackage{graphics}
\usepackage{caption}
\usepackage{tikz}
\usepackage{hyperref}
\usepackage{subcaption}

\usepackage{algorithm}
\usepackage{algpseudocode}

\usepackage{comment}

\definecolor{lightblue}{RGB}{225, 232, 239}
\usetikzlibrary{matrix,arrows.meta}

\providecommand{\keywords}[1]
{
  \small	
  {\textit{Keywords}:} #1
}

\title{\textbf{Bayesian inference for hidden Markov models under genuine multimodality with application to ecological time series}}

\author[1]{Marco A. Gallegos-Herrada\thanks{Corresponding author: marco.gallegos@mail.utoronto.ca}}
\author[1,2]{Vianey Leos-Barajas}
\author[1]{Jeffrey S. Rosenthal}

\affil[1]{Department of Statistical Sciences, University of Toronto, Canada}
\affil[2]{School of the Environment, University of Toronto, Canada}

\date{\today}

\begin{document}

\maketitle

\begin{abstract}

Bayesian inference in hidden Markov models (HMMs) can be challenging due to the presence of multimodality in the likelihood function, and consequently in the joint posterior distribution, even after correcting for label switching. The parallel tempering (PT) algorithm, a state-space augmentation method, is a widely used approach for dealing with multimodal distributions. Nevertheless, standard implementation of the PT algorithm may not always be sufficient to effectively explore the high-dimensional, complex multimodal posterior distributions that arise in HMMs. In this work, we demonstrate common pitfalls when implementing the PT algorithm for HMMs, approaches to remedy them, and introduce new non-informative prior distributions that facilitate effective posterior distribution exploration. We analyse time series of blue whale dive data with two 3-state HMMs in a Bayesian framework, one of which includes a categorical covariate in the transition probability matrix to account for the effect of sound stimuli on the whale's behavior. We demonstrate how effective implementation of the modified PT algorithm for Bayesian inference leads to effective exploration of the resultant multimodal posterior distribution and how that affects inference for the underlying movement patterns of the blue whales. 

\end{abstract}

\keywords{Bayesian inference, hidden Markov models, Markov chain Monte Carlo, parallel tempering, multimodality, time series}

\section{Introduction}

Discrete-time, finite-state hidden Markov models (HMMs) are a class of state-space models that assume that an observed time series is driven by a latent state process. They have been applied broadly across many domains, such as in ecology to connect observed movements to latent behaviors and capture-recapture data to estimate population dynamics \citep{mcclintock_uncovering_2020}, in medicine to describe the dynamics of disease progression \citep{williams_bayesian_2020}, in finance to classify market regimes such as bear and bullish periods \citep{maheu_identifying_2000} and in speech recognition \citep{juang_hidden_1991}. The natural interpretation of the hidden states as proxies for underlying behaviors allows practitioners to gain insights into unobserved processes based on observed data over time.

Model fitting can be performed, for example, via direct maximization of the likelihood function, using the HMM-specific expectation-maximization algorithm, known as the Baum-Welch algorithm \citep{baum_maximization_1970}, or in a Bayesian framework using Markov chain Monte Carlo (MCMC) methods, among other inferential approaches. When working with the (unnormalized) joint posterior distribution, one can implement general-purpose MCMC approaches such as Hamiltonian Monte Carlo or variants of the Metropolis–Hastings algorithm. Alternatively, Gibbs-type MCMC algorithms can be employed when the hidden states are estimated jointly with the parameters \citep{cappe_reversible_2003}. Regardless of the model-fitting approach, a common challenge in HMMs is the permutation invariance of hidden state labels, known as label switching, where different state labelings yield the same likelihood value. As a result, HMMs likelihood exhibit multimodality. While label switching can be mitigated by applying ordering constraints to certain parameters, multimodality beyond label switching can still persist. 

In the presence of multimodality in the likelihood function, \citep{zucchini_hidden_2017} suggest using multiple starting values when implementing numerical maximization algorithms to assess whether the same maximum is consistently identified, with the goal of finding the global maximum. However, \citep{chen_statistical_2023} describes how the presence of multiple modes affects construction of confidence intervals and subsequent coverage in a frequentist framework where different approaches, such as Wald or likelihood-based, can lead to different outcomes. In a Bayesian framework, uncertainty in parameter estimates is represented through the joint posterior distribution. If the likelihood function is multimodal and induces a multimodal posterior, inference can be conducted by constructing intervals for each local maximum, thereby accounting for uncertainty across all high-density regions. In particular, it is important that all regions of the parameter space with non-negligible posterior mass, after correcting for label switching, are thoroughly explored.

Failing to adequately account for the uncertainty associated with multiple local maxima in the joint posterior distribution can lead to biased, and potentially substantially different, parameter estimates. As such, multimodality should be treated as an inferential challenge, rather than only a computational one. A class of state-space augmentation methods has been developed to facilitate the exploration of multimodality in complex, high-dimensional settings. In particular, methods such as Parallel Tempering (PT) \citep{geyer_markov_1991} are specifically designed for this purpose, with previous attempts to apply these methods to HMMs reported in the literature \citep{sacchi_toward_2021}. However, direct application of the general PT algorithm to HMMs may not work as intended due to computational challenges, particularly in the construction of the tempered distributions used to extend the state space, typically defined as $\pi^{\beta}$ for $\beta\in(0,1)$, where $\pi$ is the target distribution, as well as in the specification of priors for model parameters. In both cases, the standard construction of the tempered replicas or an inadequate prior specification can lead in shifts in probability mass for the hotter replicas, that is, those with $\beta \rightarrow 0$, toward regions where parameter configurations correspond to near non-identifiability, resulting in poor mixing and a failure to adequately explore the original state space.

In this paper, we focus on implementation of the PT algorithm for HMMs within a Bayesian framework and how it can be effectively applied to this class of models. Specifically, we provide implementation guidelines for defining the tempered replicas and key components of the PT algorithm, together with prior specifications that mitigate shifts in probability mass towards regions associated with near non-identifiability in the hotter replicas, and thereby ensuring robust exploration of the joint posterior distribution.  We also introduce a new non-informative prior specification for the transition probability matrix entries when incorporating categorical covariates. 

This paper is organized as follows. In Section \ref{sec:background}, we introduce the mathematical formulation of HMMs, the PT algorithm and its key components, as well as the notion of genuine multimodality and the estimation of mode weights. In Section \ref{sec:methodology}, we provide implementation guidelines for applying the PT algorithm to HMMs. In Section \ref{sec:results}, we present the results of applying the PT algorithm to HMMs, following the proposed implementation guidelines, to ecological time series data, specifically blue whale dive data. In Section \ref{sec:discussion}, we summarize our work, discuss its limitations, and outline potential directions for future research.

\section{Background}\label{sec:background}

In this section, we introduce the probabilistic definition of hidden Markov models, the Parallel Tempering algorithm and its main components, as well as notions of implementation effectiveness and a formal definition of genuine multimodality. We also describe how to estimate the weights of each high-density region, or mode.

\subsection{Hidden Markov models} \label{sec:hmm}

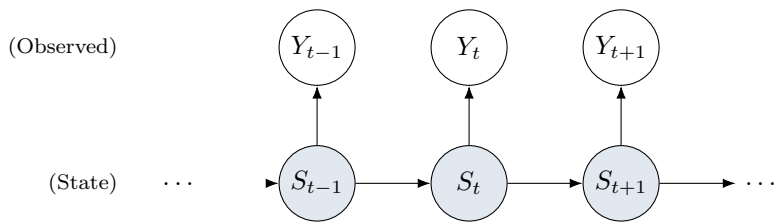
\begin{figure}[htp!]
    \centering
    \begin{tikzpicture}[node distance=2.5cm and 2.5cm,
    %every node/.style={draw,minimum size=1cm,inner sep=0pt},
    obs/.style={circle, draw, minimum size=1cm, inner sep=0pt, fill=white},
    hid/.style={circle, draw, minimum size=1cm, inner sep=0pt, fill=lightblue},
    >=stealth%,
    %thick
    ]

%\node[circle, fill=darkgray!20,   minimum size=6pt] (tNc1) at (0,0) {};

% Hidden states
\node[hid] (s1) at (0,0) {$S_{t-1}$};
\node[hid] (s2) at (2,0) {$S_{t}$};
\node[hid] (s3) at (4,0) {$S_{t+1}$};

% Observed states above
\node[obs] (y1) at (0,1.8) {$Y_{t-1}$};
%\node[obs, above=of s2] (y2) at (2,1) {$Y_{t}$};
\node[obs] (y2) at (2,1.8) {$Y_{t}$};
\node[obs] (y3) at (4,1.8) {$Y_{t+1}$};

% Labels
%\node[anchor=center] at (0, 3) {\scriptsize $t=0$};
\node[left] at (-2.5,0) {\footnotesize {(State)}};
\node[left] at (-2.5,1.8) {\footnotesize {(Observed)}};

% Hidden transitions
\draw[-{Latex}] (s1) -- (s2);
\draw[-{Latex}] (s2) -- (s3);
\node[left] at (-1.5,0) {$\ldots$};
\draw[-{Latex}] --(-1.5,0) -- (s1) ;
\draw[-{Latex}] (s3.east) -- ++(1,0) node[right] {$\ldots$};

%\draw [decorate,decoration={brace,amplitude=10pt,mirror,raise=4ex}]
  (-.5,0) -- (4.5,0) node[midway,yshift=-3em]{};

% Emissions
\foreach \i/\j in {s1/y1,s2/y2,s3/y3}
  \draw[-{Latex}] (\i) -- (\j);

\end{tikzpicture}
    \caption{Graphical representation of an HMM.}
    \label{fig:basic_hmm}
\end{figure}

A discrete-time, finite-state hidden Markov model (HMM) is a bivariate stochastic process composed of a state process $\{S_t\}_{t=1}^T$ and an observation process $\{Y\}_{t=1}^T$ \citep{zucchini_hidden_2017}. The observations $\{Y_t\}_{t=1}^T$ are taken to be conditionally independent given the states $\{S_t\}_{t=1}^T$ and assumed to be generated by a set of state-dependent distributions, $\{f(Y_t | S_t =n)\}_{n=1}^N$ for $N \in \mathbb{Z}^+$. The state process is taken to be a first-order Markov chain that evolves over time by an $N\times N$ transition probability matrix $\boldsymbol{\Gamma}$ with entries $\gamma_{ij} = \Pr(S_t=j|S_{t-1}=i)$ for $i,j \in \{1, \ldots, N\}$. Finally, the initial state distribution $\boldsymbol{\delta}$ has entries $\delta_n = \Pr(S_1 = n)$ for $n\in \{1, \ldots, N\}$. See figure \ref{fig:basic_hmm} for a graphical representation of the structure of a basic HMM. 

\noindent The likelihood of an HMM can be written as a matrix product of the initial state distribution vector $\boldsymbol{\delta}$, transition matrix $\boldsymbol{\Gamma}$ and state-dependent distributions,
\begin{equation} \label{eq:likelihood}
L_T = \boldsymbol{\delta}^{\top} \boldsymbol{D}(y_1) \boldsymbol{\Gamma} \boldsymbol{D}(y_2) \cdots \boldsymbol{\Gamma} \boldsymbol{D}(y_T)\boldsymbol{1}_N,
\end{equation}
where $D(y_t)$ is a diagonal matrix with entries $f(y_t \mid s_t)$, for $t\in \{1,\ldots , T\}$, and $\boldsymbol{1}_N$ is an $N$-dimensional vector with 1 entries. Given the recursive nature of the likelihood, it can be effectively evaluated with the forward algorithm \citep{zucchini_hidden_2017}. 

HMMs can accommodate multivariate observation processes and incorporation of covariates in the transition probability matrix, among other possible extensions. For a $P$-dimensional multivariate observation process, $\{\boldsymbol{Y}_t\}_{t=1}^T$, we assume conditional contemporaneous independence such that $f(\boldsymbol{Y}_t | S_t) = \prod_{p=1}^P f(Y_{tp}|S_t)$. HMMs can also be extended to incorporate covariates in the transition probability matrix via a multinomial logit link function, 
\begin{equation} \label{eq:tpm_cov}
    \gamma_{ij}(z_t) = \dfrac{\exp\left(\eta_{ij}(z_t)\right)}{\sum_{l=1}^n\exp(\eta_{il}(z_t))} \quad \quad \quad \eta_{ij}(z_t) = 
    \begin{cases}
        0 & \text{if } i=j\\
        \alpha_0^{(ij)} + \alpha_1^{(ij)}z_t & \text{if } i\neq j
    \end{cases}.
\end{equation}

\subsection{Parallel Tempering}\label{subsec:pt}

A popular algorithm used to explore multimodal probability distribution functions is the parallel tempering (PT) algorithm, also known as replica exchange or Metropolis-coupled Markov chain Monte Carlo, independently introduced in statistics \citep{geyer_markov_1991} and physics \citep{hukushima_exchange_1996}, and has been successfully applied in the fields of biology \citep{muller_adaptive_2019}, chemistry \citep{lin_parallel_2003}, physics \citep{diaz_where_2020} and machine learning \citep{chandra_langevin-gradient_2018}, \citep{desjardins_deep_2014}. See also \citep{swendsen_replica_1986} for an earlier related proposal. Parallel tempering is a state-space augmentation method that extends the parameter space of the target distribution $\pi$ using auxiliary distributions or tempered replicas $\pi_{\beta}$, where in a Bayesian framework $\pi$ is the joint posterior distribution. The tempered replicas are commonly defined as power-tempered versions of target distribution, 
\begin{equation}
\pi_{\beta}(\boldsymbol{x})\propto [\pi(\boldsymbol{x})]^{\beta}.
\end{equation} 
These power transformations make the modes of $\pi$ less separated as the power values decrease, resulting in high-density regions that are closer together. The tempered replicas are indexed by a sequence of inverse temperatures $0\leq \beta_M < \beta_{M-1} < \ldots < \beta_1 < \beta_0 = 1$, referred to as the inverse temperature schedule or inverse temperature ladder. The coldest replica $\pi_{\beta_0}$ is defined to be the original distribution of interest $\pi$, whereas the hottest replica $\pi_{\beta_n}$ is a transformation of the distribution $\pi$ that can be fully explored using standard MCMC approaches. Nonetheless, depending on the structure of the distribution, tempering can be performed in different ways. For example, in a Bayesian problems, it is common practice to temper only the likelihood component, so the sequence of tempered replicas interpolates prior and posterior. Many other
schemes are possible; see \citep{geyer_annealing_1995}, \citep{paquet_perturbation_2009}, \citep{cameron_recursive_2014},\citep{surjanovic_parallel_2023}, for alternative constructions of tempered replicas. 

The PT algorithm runs a Markov chain on the augmented state space $\mathcal{X}^{M+1}$ with the invariant distribution
\[
\pi_M(\boldsymbol{x_0},\boldsymbol{x_1},\ldots,\boldsymbol{x_M}) \propto \pi_{\beta_0}(\boldsymbol{x_0})\pi_{\beta_1}(\boldsymbol{x_1})\cdots\pi_{\beta_M}(\boldsymbol{x_M}).
\]
The Markov chain targeting $\pi_M$ alternates between two Markovian moves: within-temperature moves and swap moves. During the within-temperature moves, each $\boldsymbol{x_i}$ is updated $U$ times, where $\boldsymbol{x_i}$ is the state from the $i$-th chain targeting $\pi_{\beta_i}$. Performing multiple within-temperature moves between each temperature swap proposal, that is, $U>1$, may improve efficiency; \citep{roberths_optimal_2026} argues that $O(\sqrt{d})$ such moves are optimal in some contexts, where $d$ is the dimension of the state space $\mathcal{X}$. For the swap moves, there are two main schemes: stochastic even-odd (SEO) swaps and deterministic even-odd (DEO) swaps. In the SEO scheme, referred to as reversible PT, a pair of inverse temperatures $\beta_k$ and $\beta_{k+1}$ is chosen uniformly from all the possible pairs of adjacent inverse temperatures, and the swap proposal is constructed by swapping the corresponding chain states $\boldsymbol{x_k}$, $\boldsymbol{x_{k+1}}$,
\[
(\boldsymbol{x_1},\ldots,\boldsymbol{x_{k+1}},\boldsymbol{x_k},\ldots,\boldsymbol{x_M}),
\]
and the swap move is accepted with probability
\[
A_{k}= 1\wedge \dfrac{\pi_{\beta_{k+1}}(\boldsymbol{x_k}) \pi_{\beta_{k}}(\boldsymbol{x_{k+1}}) }{\pi_{\beta_k}(\boldsymbol{x_k}) \pi_{\beta_{k+1}}(\boldsymbol{x_{k+1}})}.
\]
The DEO scheme, introduced by \citep{okabe_replica-exchange_2001} and referred to as non-reversible PT, differs from uniform selection of adjacent inverse-temperature pairs. Instead, it deterministically alternates between even and odd swap moves after within-temperature steps. See \citep{syed_non-reversible_2022} for further details.

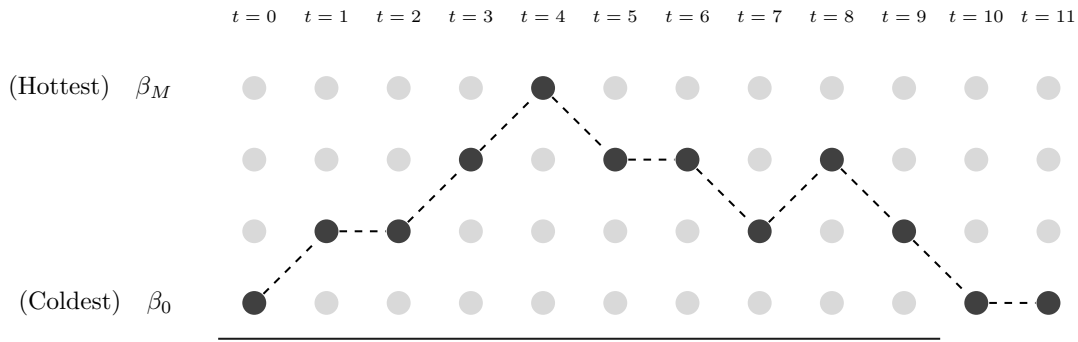
\begin{figure}[htp!]
    \centering
        \scalebox{0.95}{
\begin{tikzpicture}[>=stealth,thick]

% ----------------------------------------------------------------
% We have 5 rows: t0, t1, t2, t3, tN
% Each row has 4 columns of dots (columns at x=0, x=2, x=4, x=6).
% We'll label the rows on the left side.
% We'll draw arcs from column i to column i+1 in each row.
% ----------------------------------------------------------------

% Y-coordinates for each row
% Row t0 at y=0, t1 at y=-1, t2 at y=-2, t3 at y=-3, tN at y=-4
% We'll place the text label at x=-1 for each row.

\node[anchor=center] at (0, 1) {\scriptsize $t=0$};
\node[anchor=center] at (1, 1) {\scriptsize $t=1$};
\node[anchor=center] at (2, 1) {\scriptsize $t=2$};
\node[anchor=center] at (3, 1) {\scriptsize $t=3$};
\node[anchor=center] at (4, 1) {\scriptsize $t=4$};
\node[anchor=center] at (5, 1) {\scriptsize $t=5$};
\node[anchor=center] at (6, 1) {\scriptsize $t=6$};
\node[anchor=center] at (7, 1) {\scriptsize $t=7$};
\node[anchor=center] at (8, 1) {\scriptsize $t=8$};
\node[anchor=center] at (9, 1) {\scriptsize $t=9$};
\node[anchor=center] at (10, 1) {\scriptsize $t=10$};
\node[anchor=center] at (11, 1) {\scriptsize $t=11$};

% 1) Row t0
\definecolor{airforceblue}{rgb}{0.36,0.54,0.66}
\definecolor{amethyst}{rgb}{0.6,0.4,0.8}
\definecolor{amaranth}{rgb}{0.9,0.17,0.31}
\definecolor{applegreen}{rgb}{0.55,0.71,0.0}
\node[anchor=east] at (-1, 0) {$\text{(Hottest)\quad} \beta_M$};
\node[circle, fill=darkgray!20,   minimum size=6pt] (tNc1) at (0,0) {};
\node[circle, fill=darkgray!20,   minimum size=6pt] (tNc2) at (1,0) {};
\node[circle, fill=darkgray!20, minimum size=6pt] (tNc3) at   (2,0) {};
\node[circle, fill=darkgray!20, minimum size=6pt] (tNc4) at   (3,0) {};
\node[circle, fill=darkgray,   minimum size=6pt] (tNc5) at (4,0) {};
\node[circle, fill=darkgray!20,   minimum size=6pt] (tNc6) at (5,0) {};
\node[circle, fill=darkgray!20,   minimum size=6pt] (tNc7) at (6,0) {};
\node[circle, fill=darkgray!20,   minimum size=6pt] (tNc8) at (7,0) {};
\node[circle, fill=darkgray!20,   minimum size=6pt] (tNc9) at (8,0) {};
\node[circle, fill=darkgray!20,   minimum size=6pt] (tNc10) at (9,0) {};
\node[circle, fill=darkgray!20,   minimum size=6pt] (tNc11) at (10,0) {};
\node[circle, fill=darkgray!20,   minimum size=6pt] (tNc12) at (11,0) {};

% 3) Row t2
\node[anchor=east] at (-1, -1) {};%{$\beta_2$};
\node[circle, fill=darkgray!20, minimum size=6pt] (t2c1) at (0,-1) {};
\node[circle, fill=darkgray!20, minimum size=6pt] (t2c2) at (1,-1) {};
\node[circle, fill=darkgray!20, minimum size=6pt] (t2c3) at (2,-1) {};
\node[circle, fill=darkgray, minimum size=6pt] (t2c4) at    (3,-1) {};
\node[circle, fill=darkgray!20,   minimum size=6pt] (t2c5) at (4,-1) {};
\node[circle, fill=darkgray,   minimum size=6pt] (t2c6) at (5,-1) {};
\node[circle, fill=darkgray,   minimum size=6pt] (t2c7) at (6,-1) {};
\node[circle, fill=darkgray!20,   minimum size=6pt] (t2c8) at (7,-1) {};
\node[circle, fill=darkgray,   minimum size=6pt] (t2c9) at (8,-1) {};
\node[circle, fill=darkgray!20,   minimum size=6pt] (t2c10) at (9,-1) {};
\node[circle, fill=darkgray!20,   minimum size=6pt] (t2c11) at (10,-1) {};
\node[circle, fill=darkgray!20,   minimum size=6pt] (t2c12) at (11,-1) {};

% 4) Row t1
\node[anchor=east] at (-1, -2) {};%{$\beta_1$};
\node[circle, fill=darkgray!20, minimum size=6pt] (t1c1) at (0,-2) {};
\node[circle, fill=darkgray, minimum size=6pt] (t1c2) at    (1,-2) {};
\node[circle, fill=darkgray,   minimum size=6pt] (t1c3) at  (2,-2) {};
\node[circle, fill=darkgray!20, minimum size=6pt] (t1c4) at (3,-2) {};
\node[circle, fill=darkgray!20,   minimum size=6pt] (t1c5) at (4,-2) {};
\node[circle, fill=darkgray!20,   minimum size=6pt] (t1c6) at (5,-2) {};
\node[circle, fill=darkgray!20,   minimum size=6pt] (t1c7) at (6,-2) {};
\node[circle, fill=darkgray,   minimum size=6pt] (t1c8) at (7,-2) {};
\node[circle, fill=darkgray!20,   minimum size=6pt] (t1c9) at (8,-2) {};
\node[circle, fill=darkgray,   minimum size=6pt] (t1c10) at (9,-2) {};
\node[circle, fill=darkgray!20,   minimum size=6pt] (t1c11) at (10,-2) {};
\node[circle, fill=darkgray!20,   minimum size=6pt] (t1c12) at (11,-2) {};

% 5) Row t0
\node[anchor=east] at (-1, -3) {$\text{(Coldest)\quad} \beta_0$};
\node[circle, fill=darkgray,   minimum size=6pt] (t0c1) at (0,-3) {};
\node[circle, fill=darkgray!20,   minimum size=6pt] (t0c2) at (1,-3) {};
\node[circle, fill=darkgray!20,   minimum size=6pt] (t0c3) at (2,-3) {};
\node[circle, fill=darkgray!20,   minimum size=6pt] (t0c4) at (3,-3) {};
\node[circle, fill=darkgray!20,   minimum size=6pt] (t0c5) at (4,-3) {};
\node[circle, fill=darkgray!20,   minimum size=6pt] (t0c6) at (5,-3) {};
\node[circle, fill=darkgray!20,   minimum size=6pt] (t0c7) at (6,-3) {};
\node[circle, fill=darkgray!20,   minimum size=6pt] (t0c8) at (7,-3) {};
\node[circle, fill=darkgray!20,   minimum size=6pt] (t0c9) at (8,-3) {};
\node[circle, fill=darkgray!20,   minimum size=6pt] (t0c10) at (9,-3) {};
\node[circle, fill=darkgray,   minimum size=6pt] (t0c11) at (10,-3) {};
\node[circle, fill=darkgray,   minimum size=6pt] (t0c12) at (11,-3) {};

% Swapping proposal for each dot column
\draw[dashed,-] (t0c1) to (t1c2);
\draw[dashed,-] (t1c2) to (t1c3);
\draw[dashed,-] (t1c3) to (t2c4);
\draw[dashed,-] (t2c4) to (tNc5);
\draw[dashed,-] (tNc5) to (t2c6);
\draw[dashed,-] (t2c6) to (t2c7);
\draw[dashed,-] (t2c7) to (t1c8);
\draw[dashed,-] (t1c8) to (t2c9);
\draw[dashed,-] (t2c9) to (t1c10);
\draw[dashed,-] (t1c10) to (t0c11);
\draw[dashed,-] (t0c11) to (t0c12);

\draw[-,color=black!90] (-.5,-3.5) to (9.5,-3.5);

\end{tikzpicture}
}

    \caption{Illustration of a round trip occurrence at time $t = 10$ with $M = 3$ and $U = 1$ within-temperature updates prior to attempting a swap move.}
    \label{fig:round_trip}
\end{figure}

Tracking the exchange of chain states during swap moves serves as a metric for assessing the performance of the PT algorithm. The completion of round trips, that is, information from the coldest replica reaching the hottest replica and returning back to the coldest, indicates how effectively the information is being passed on. See figure \ref{fig:round_trip} for an illustration of the occurrence of a round trip, when using $M=3$. For reversible PT, minimizing the expected round trip time, this is, the number of iterations it takes on average to complete a round trip, has been shown to be equivalent to optimizing swap acceptance rates under some regularity conditions \citep{nadler_generalized_2007}.

\subsubsection*{Methods for setting the inverse temperature schedule}

The number of round trips strongly depends on the choice of inverse temperature levels $\beta_0,\ldots, \beta_M$. Several methods have been proposed in the literature for constructing the temperature schedule. A common approach is a geometric schedule \citep{kofke_acceptance_2002}, consisting of a geometric progression between the coldest ($\beta_0$) and hottest ($\beta_M$) inverse temperatures. In this case, the inverse temperatures are defined as $\beta_m = \beta_0 R^{m}$, where $R = (\beta_M / \beta_0)^{1/M}$.

Other approaches aim to achieve a uniform swap acceptance rate across tempered replicas. \citep{predescu_incomplete_2004} showed that a geometric schedule aligns with a uniform swap acceptance rate when the heat capacity is constant across replicas. \citep{nadler_generalized_2007} demonstrated that constant swap acceptance rates are part of the optimal solution when maximizing flow current under certain regularity conditions. Similarly, \citep{atchade_towards_2011}, \citep{roberts_minimising_2014}, showed that, under strong regularity conditions on the target distribution, the optimal constant swap acceptance rate between adjacent inverse temperatures is $A^{*} = 0.234$. Consistently, \citep{kone_selection_2005} and \citep{lingenheil_efficiency_2009} obtained the same optimal rate when optimizing the diffusion of a tempered replica along the inverse temperature schedule.

Other methods focus on constructing an inverse temperature schedule that concentrates around simulation bottlenecks. \citep{katzgraber_feedback-optimized_2006} proposed optimizing the current flow by identifying bottlenecks through measurements of local diffusivity of tempered replicas and placing temperature values more densely in those regions. Inverse-linear schedules have also been used in practice. In this approach, inverse temperatures are defined as $\beta_m = \beta_0 + (\beta_M - \beta_0)(m/M)$. \citep{rozada_effects_2019} showed that this strategy performs well for sparse spin-glass problems.

\subsubsection*{Genuine multimodality}\label{subsec:genuine_multimodality}

The invariance to relabeling the states from the likelihood in HMMs induces a system of equivalence classes $\mathcal{P}$ on the parameter space $\Theta$. Two elements $\boldsymbol{\theta_1}, \boldsymbol{\theta_2}$ of $\Theta$ result in belonging to the same equivalence class if there exists a permutation of the hidden state labels such that, after reordering the elements according to that permutation, they are equal. Formally,
\[
\boldsymbol{\theta_1} \sim \boldsymbol{\theta_2} \iff \exists\nu \in \mathrm{Perm}(S) \text{ such that } \boldsymbol{\theta_1} = \nu(\boldsymbol{\theta_2}),
\]
where Perm($S$) denotes the set of all possible permutations of $S$ items. Assigning exchangeable priors to the initial state distribution, the rows of the transition probability matrix, and the state-dependent parameters extends the relabeling invariance of the likelihood to the posterior distribution \citep{fruhwirth-schnatter_markov_2001}. Consequently, the posterior distribution $p(\cdot \mid y)$ of an $N$-state HMM will, at least theoretically, exhibit $N!$ symmetric modes, where a mode is defined as a local maximum in $p(\cdot \mid y)$. \cite{grun_dealing_2009} defines $p(\cdot \mid y)$ to be genuinely multimodal if
\[
\exists\,\boldsymbol{\theta_1}, \boldsymbol{\theta_2} \in\mathcal{M} \text{ such that } \boldsymbol{\theta_1} \neq \nu(\boldsymbol{\theta_2}) \quad \forall \text{ Perm}(S),
\]
with $\mathcal{M}$ the set of modes of $p(\cdot \mid y)$. Let $\hat{\Theta} = \text{ident}(\Theta) \subseteq \Theta$ be the subset of parameterizations containing only one permutation from each equivalence class, and let $\hat{\mathcal{M}}\subseteq \mathcal{M}$ denote the subset of modes located in $\hat{\Theta}$. To conduct robust inference, it is crucial to account for the case $|\hat{\mathcal{M}}| > 1$, indicating there is more than one genuine mode. When multiple genuine modes are present in the posterior parameter space, their exploration via the PT algorithm, or any other MCMC method, must respect the corresponding mode weights, i.e. the probability mass associated with each mode. If mode $h \in \hat{\mathcal{M}}$ is well separated from the others such that it is located in a subset $\hat{\Theta}_h \subseteq \hat{\Theta}$ containing only mode $h$ and no other genuine mode, then the weight $w_h$ can be approximated by estimating $\mathbb{P}_{p(\cdot \mid y)}(\boldsymbol{\theta} \in \hat{\Theta}_h)$. Specifically, given $K$  samples from the coldest chain, with the first $B$ samples discarded as burn-in, the running weight estimate $\hat{w}_{h,K}$ for mode $h$ at iteration $K$ is computed using the samples $\boldsymbol{\theta^{(B)}}, \boldsymbol{{\theta^{(B+1)}}}, \ldots,\boldsymbol{ \theta^{(K)}}$,
\[
\hat{w}_{h,K} = \dfrac{1}{N-B + 1}\sum_{i=B}^{N} \boldsymbol{1}_{\{\boldsymbol{\theta^{(i)}} \in \hat{\Theta}_h \}}(\boldsymbol{\theta^{(i)}}).
\]
In the case that the subset $\hat{\Theta}_h$ can be defined by thresholds for only one of its dimensions, this is, $\hat{\Theta}_h = \{ \boldsymbol{\theta} = (\theta_1,\ldots, \theta_V)  : L < \theta_s < U \}$, then $\hat{w}_{h,K}$ can be simply computed by
\[
\hat{w}_{h,K} = \dfrac{1}{N-B + 1}\sum_{i=B}^{N} \boldsymbol{1}_{\{L < \theta_s^{(i)} < U\}}(\theta_{s}^{(i)}).
\]

\section{Methodology}\label{sec:methodology}

In this section, we introduce novel prior specifications for the unconstrained parameters arising from the transformation of the transition probabilities in Equation \ref{eq:tpm_cov}, which induce equal weighting on the simplex of the transition probability rows. We also propose a prior formulation for the coefficients associated with incorporating a categorical covariate into the transition probabilities that similarly ensures equal weighting on the simplex. We then detail the implementation of the PT algorithm for HMMs, including the tempering of the replicas, the construction of the inverse temperature schedule $\{\beta_0, \beta_1, \ldots, \beta_M\}$, and practical considerations for implementing the PT algorithm.

\subsection{Bayesian framework for HMMs}\label{subsec:bayes_hmm}

A full specification for HMMs in a Bayesian framework requires assigning prior distributions for all parameters in the likelihood function given in Equation \ref{eq:likelihood}. Setting prior distributions for the parameters of the state-dependent distributions depends on the forms of $f(\cdot)$. A common practice when specifying weakly informative priors is to use exchangeable priors on the state-dependent distribution parameters across all hidden states; see \citep{fruhwirth-schnatter_markov_2001} for a detailed discussion. For the initial state distribution vector $\boldsymbol{\delta}$, a common choice of non-informative prior is a Dirichlet distribution with parameter vector $ \boldsymbol{1}_N$. Similarly, for the entries of a time-homogeneous $\boldsymbol{\Gamma}$, each row is given prior $\boldsymbol{\gamma_{i\cdot}} \sim \text{Dirichlet}(\boldsymbol{1}_N)$, for $i \in \{1, \ldots, N\}$. However, this prior cannot be used directly when the transition probabilities are reparameterized into unconstrained components, such as when incorporating covariates in Equation \ref{eq:tpm_cov}, and it is not straightforward how priors for $\boldsymbol{\alpha_{0}^{(ij)}}, \boldsymbol{\alpha_{1} ^{(ij)}}$ affect the behavior of $\boldsymbol{\gamma_{i\cdot} }$.

\begin{figure}[htp!]  
%\vspace*{-0mm}
\centering
\begin{subfigure}{.45\textwidth}
  %\centering
  \includegraphics[trim={0 0 0 2cm},clip,width=1\linewidth]{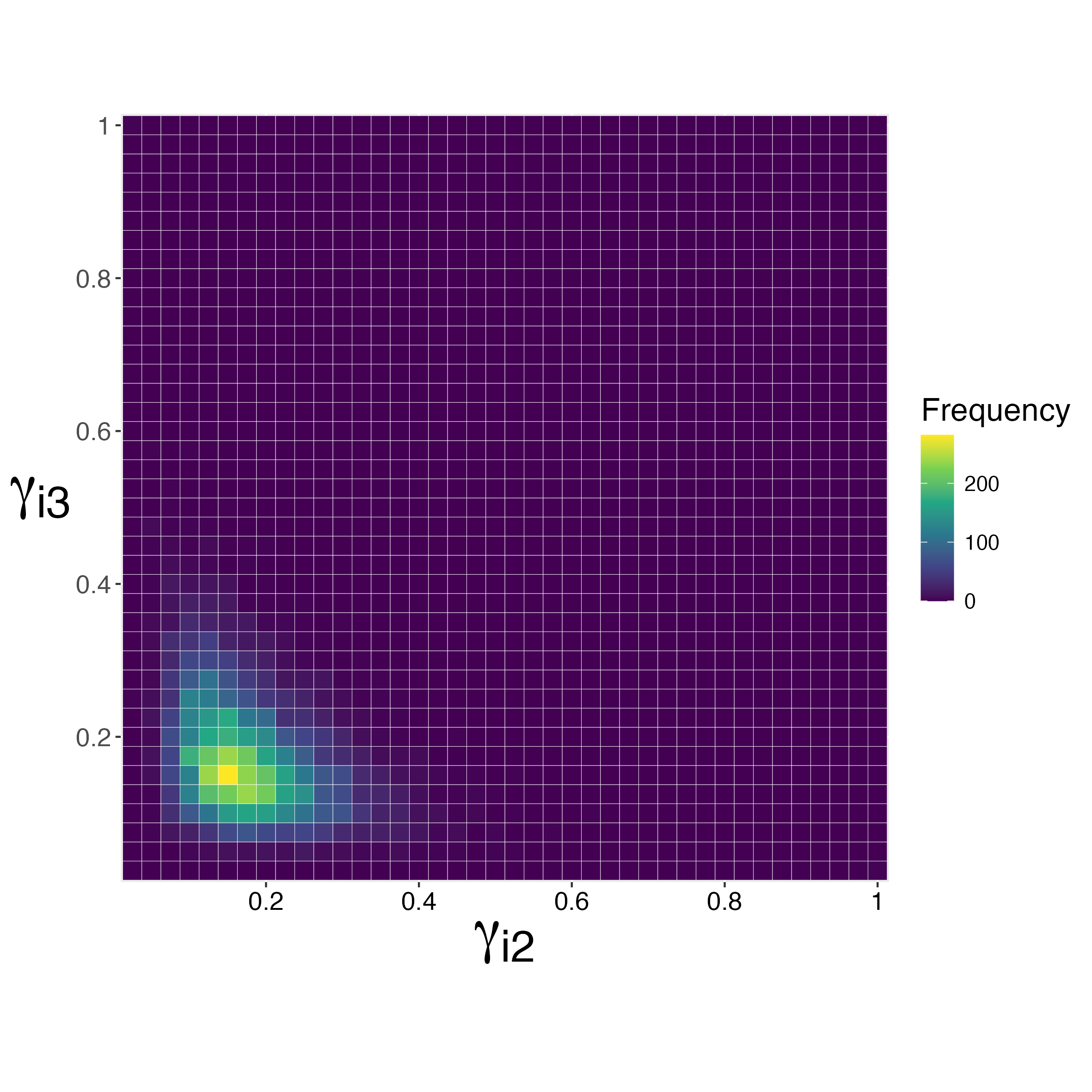}
  \vspace*{-12mm}
  \caption{\centering $\alpha_{0}^{(ij)} \sim \mathcal{N}(0,\sigma = 0.5)$}
\end{subfigure}%
\begin{subfigure}{.45\textwidth}
  \centering
  \includegraphics[trim={0 0 0 2cm},clip,width=1\linewidth]{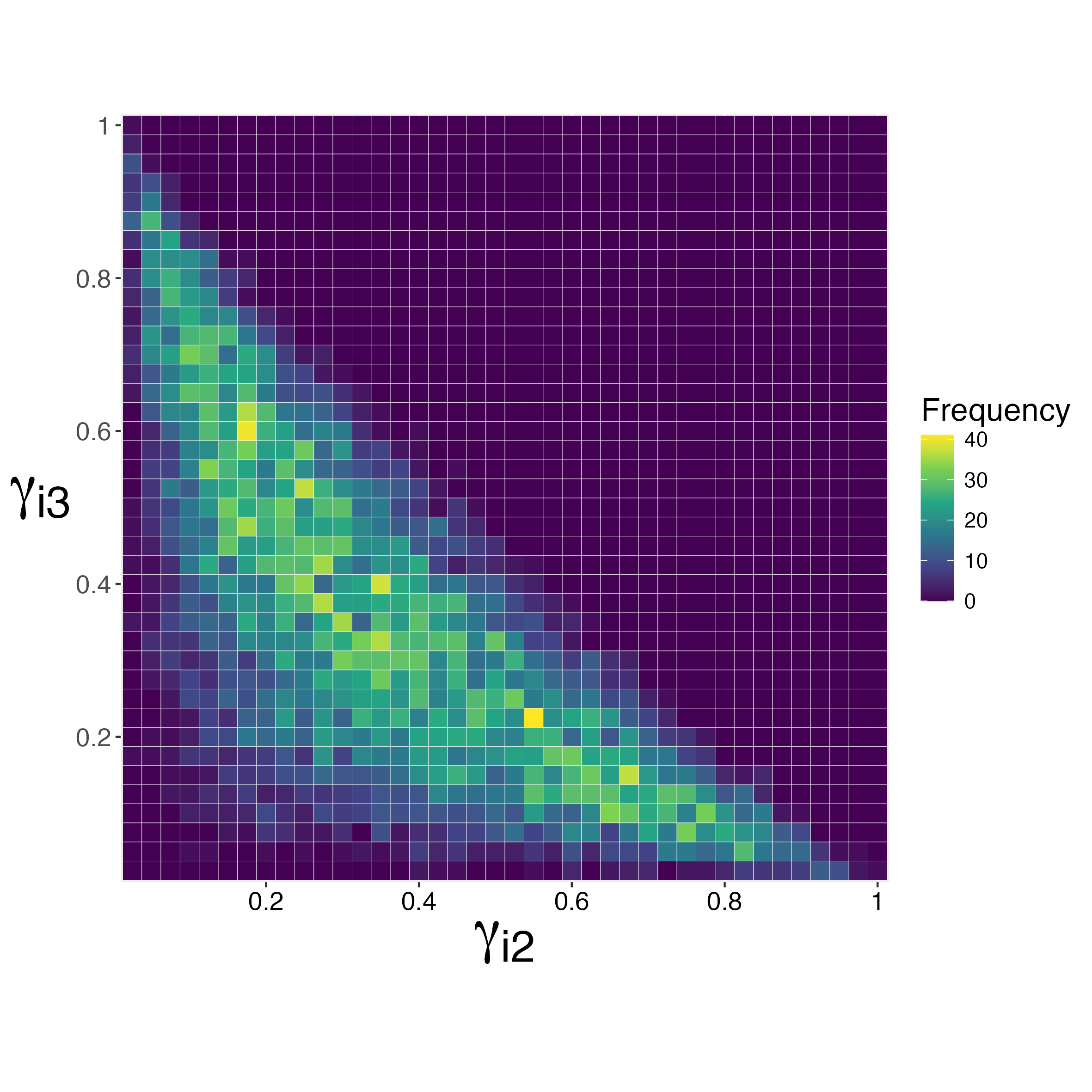}
  \vspace*{-12mm}
  \caption{\centering $\alpha_{0}^{(ij)} \sim \mathcal{N}(0,\sigma = 1)$}
\end{subfigure}%

\begin{subfigure}{.45\textwidth}
  \centering    
  \vspace*{5mm}
  \includegraphics[trim={0 0 0 2cm},clip,width=1\linewidth]{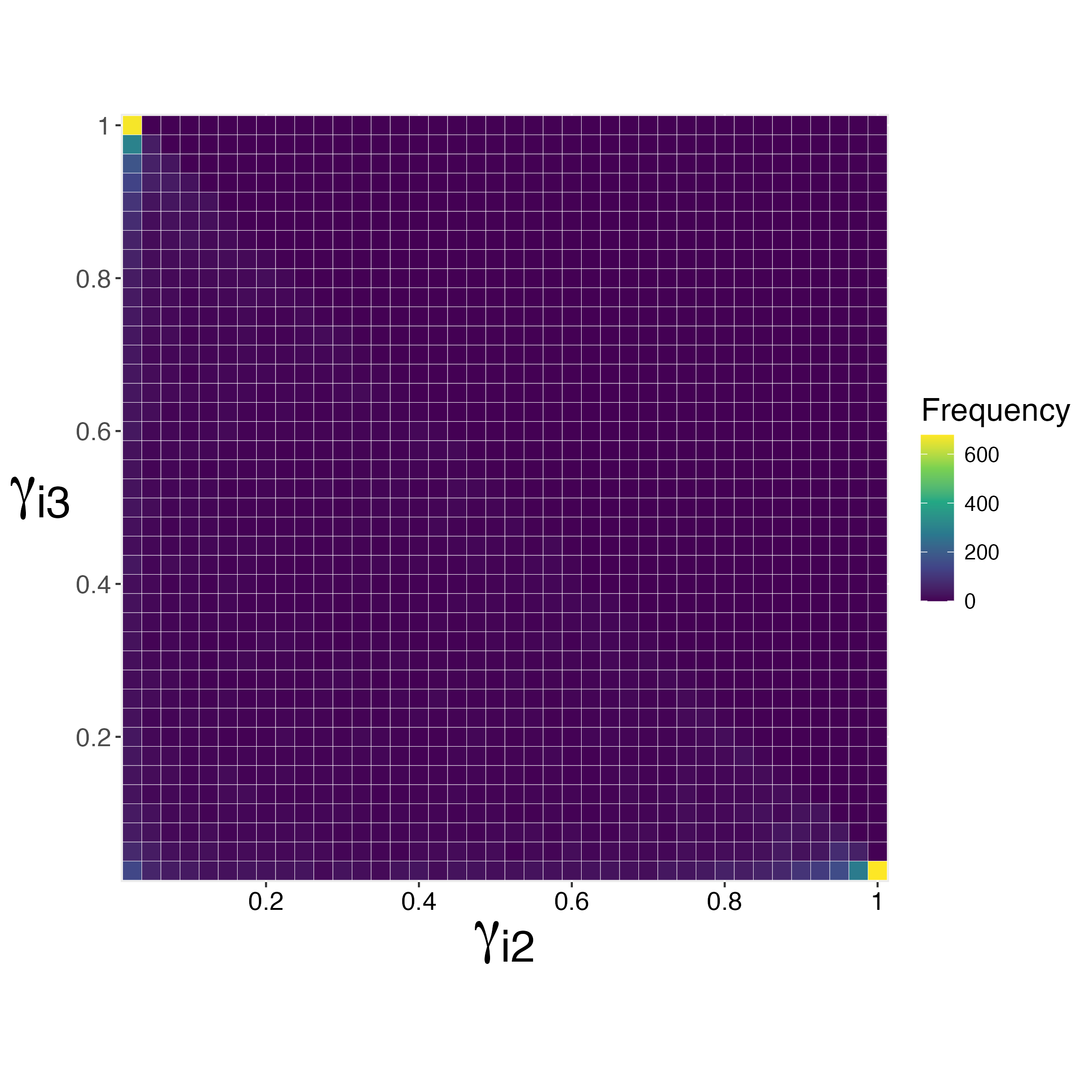}
  \vspace*{-12mm}
  \caption{\centering $\alpha_{0}^{(ij)} \sim \mathcal{N}(0,\sigma = 3)$}
\end{subfigure}
\begin{subfigure}{.45\textwidth}
  \centering
    \vspace*{5mm}
  \includegraphics[trim={0 0 0 2cm},clip,width=1\linewidth]{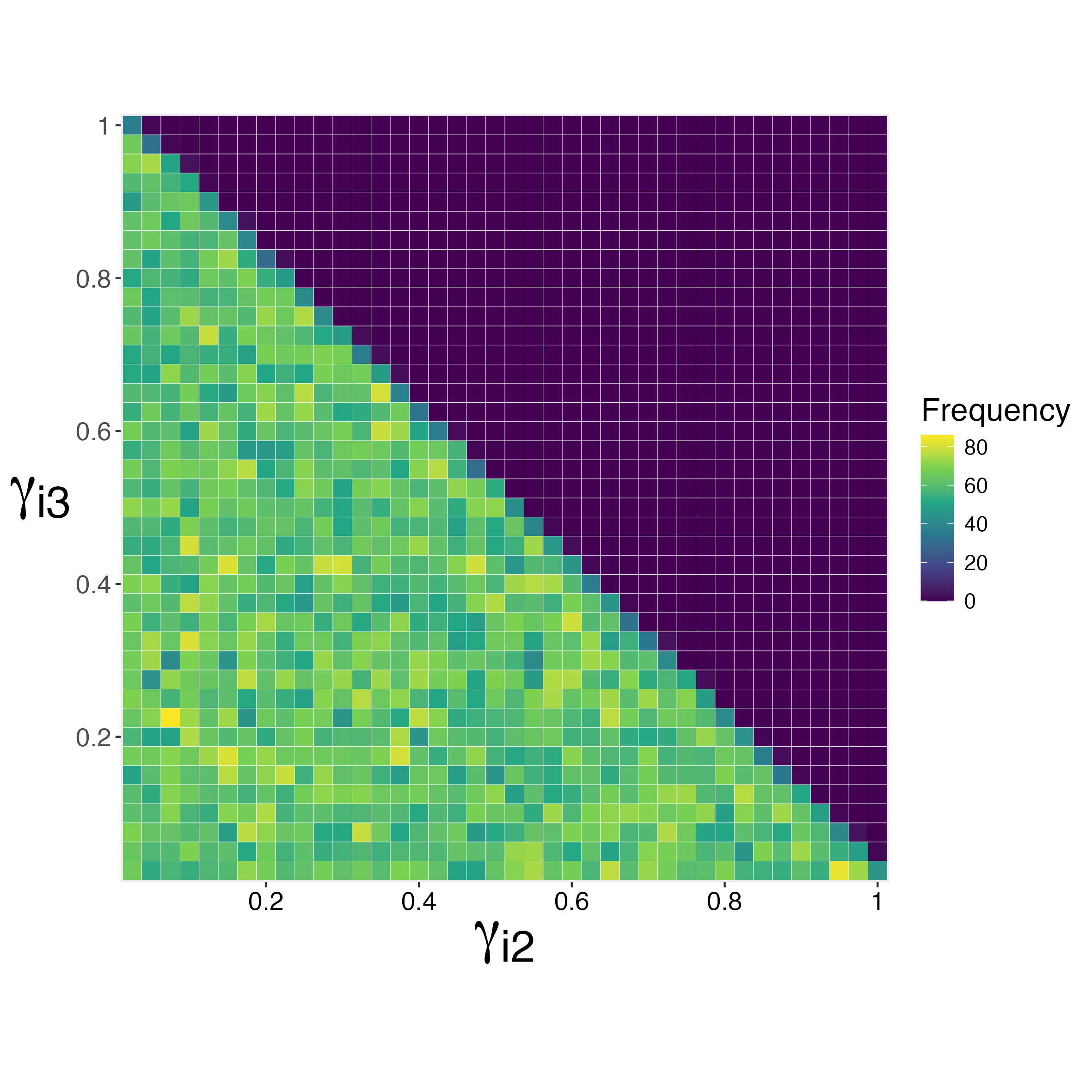}
  \vspace*{-12mm}
  \caption{\centering  $(\gamma_{i1}, \gamma_{i2}, \gamma_{i3} ) \sim \text{Dir}(\boldsymbol{1}_N) $}
\end{subfigure}
    \caption{Joint distribution of $(\gamma_{i2},\gamma_{i3})$ induced by different choices of priors for the working parameters $\alpha_{0}^{(ij)}$.}
    \label{fig:induced_transition_probs}
\end{figure}

For the individual components of the vector $\boldsymbol{\alpha_{0}^{(ij)}}$ a normal distribution $\mathcal{N}\left(\mu_{\alpha_{0}^{(ij)}}, \sigma_{\alpha_{0} ^{(ij)}}\right)$ can be used. However, specifying $\mu_{\alpha_{0}^{(ij)}}$ and $\sigma_{\alpha_{0}^{(ij)}}$ to reflect non-informative or weakly informative priors on the scale of the unconstrained parameters can induce highly skewed, informative distributions for $\boldsymbol{\gamma_{i\cdot}}$ a priori. See figure \ref{fig:induced_transition_probs} for a visualization of the joint distribution corresponding to two elements of the $i$-th row $\boldsymbol{\gamma_{i\cdot}} = (\gamma_{i1}, \ldots, \gamma_{iN})$ induced by normal priors when $N=3$. 

We introduce a prior distribution for the unconstrained baseline components $\alpha_{0}^{(ij)}$ that induces a $\text{Dirichlet}(\boldsymbol{1}_N)$ distribution on $\boldsymbol{\gamma_{i\cdot}}$. Assume the following:
\begin{align*}
-\alpha_{0}^{(ij)}\mid \zeta_{ii} &\sim \text{Gumbel}(\zeta_{ii},  1),\\
 -\zeta_{ii} &\sim \text{Gumbel}(0,1),
\end{align*}
where $\zeta_{ii}$ is a latent variable for the $i$-th row $\boldsymbol{\gamma}_{i\cdot}$. This leads to $(\gamma_{i1}, \ldots, \gamma_{iN}) \sim \text{Dirichlet}(\boldsymbol{1}_N)$, for $i \in \{1, \ldots, N\}$. For the complete derivation of the proposed prior distributions and how they induce  a Dirichlet distribution, see Appendix \ref{appdx:priors_derivation}. 

A similar approach can be used to induce a non-informative distribution on the transition probability rows for the coefficients resulting from incorporating a categorical covariate in the transition probability matrix, as shown in Equation \ref{eq:tpm_cov}. In particular, when the covariate is categorical, the formulation in Equation \ref{eq:tpm_cov}, originally introduced for a continuous covariate, is modified with the inclusion of auxiliary variables to avoid model non-identifiability. In this case, being $c_0, \ldots, c_L$ the possible categories of $z_t$, the linear formulation becomes
\begin{equation}\label{eq:tpm_cov_categorical}
\eta_{ij}(z_t) = \alpha_{0}^{(ij)} + \sum_{l=1}^{L} \alpha_{1l}^{(ij)} , \mathbf{1}_{\{z_t = c_l\}},
\quad \text{for } i \neq j.    
\end{equation}
To construct a prior for the coefficient related to the covariate
that induces uniform weights to the transition probability rows, we leverage the prior setting defined for the baseline components. Specifically, given the construction of the baseline components $\alpha_{0}^{(ij)}$ with the auxiliary variables $\zeta_{ii}$, the following prior assumption induces equal weights for the transition probability rows for any outcome of $z_t$:
\begin{align*}
    -\alpha_{1l}^{(ij)}\mid \alpha_{0}^{(ij)}, \zeta_{ii} &\sim \text{Gumbel}(\alpha_{0}^{(ij)} + \zeta_{ii}, 1).
\end{align*}
Additionally, we assume the $\alpha_{1l}^{(ij)}$ coefficients to be independent conditional on $\alpha_{0}^{(ij)}, \zeta_{ii}$. The dependence on the auxiliary variables $\zeta_{ii}$ guarantees identifiability, and from the prior formulation it follows that $(\gamma_{i1}(z_t), \ldots, \gamma_{iN}(z_t)) \sim \text{Dirichlet}(\boldsymbol{1}_N)$ for any outcome of $z_t$, which can be interpreted as no additional information is introduced about whether the treatment or latent variable affects the entries of the transition probability matrix compared to the baseline. For the detailed derivation of this prior setting, see Appendix \ref{appdx:priors_derivation}.

\subsection{Parallel tempering for hidden Markov models}

\begin{figure}[htp!]  
%\vspace*{-0mm}
\centering
\begin{subfigure}{.33\textwidth}
  %\centering
  \includegraphics[trim={0 0 0 2cm},clip,width=1\linewidth]{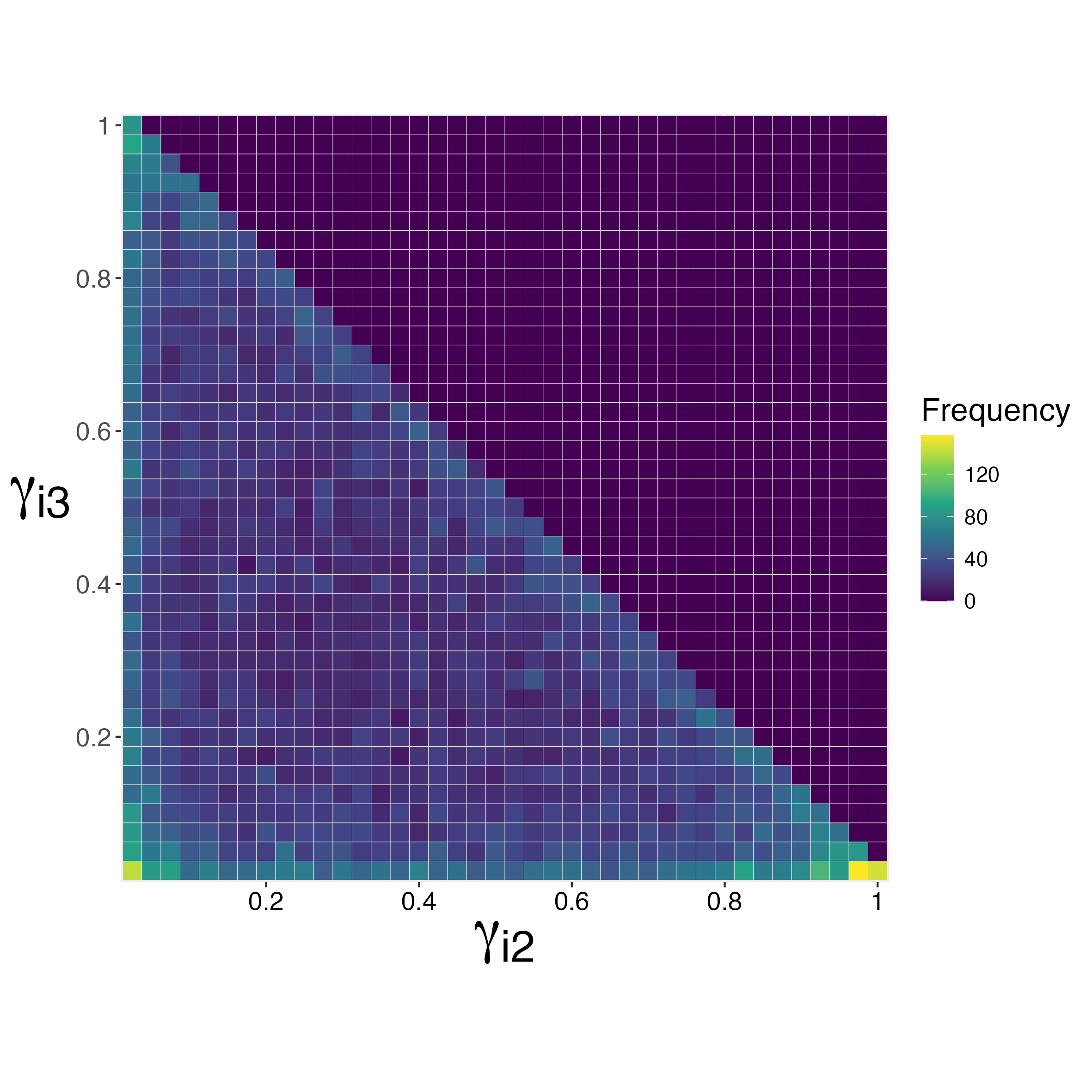}
  \vspace*{-8mm}
  \caption{\centering $\beta = 0.667$}
\end{subfigure}%
\begin{subfigure}{.33\textwidth}
  \centering
  \includegraphics[trim={0 0 0 2cm},clip,width=1\linewidth]{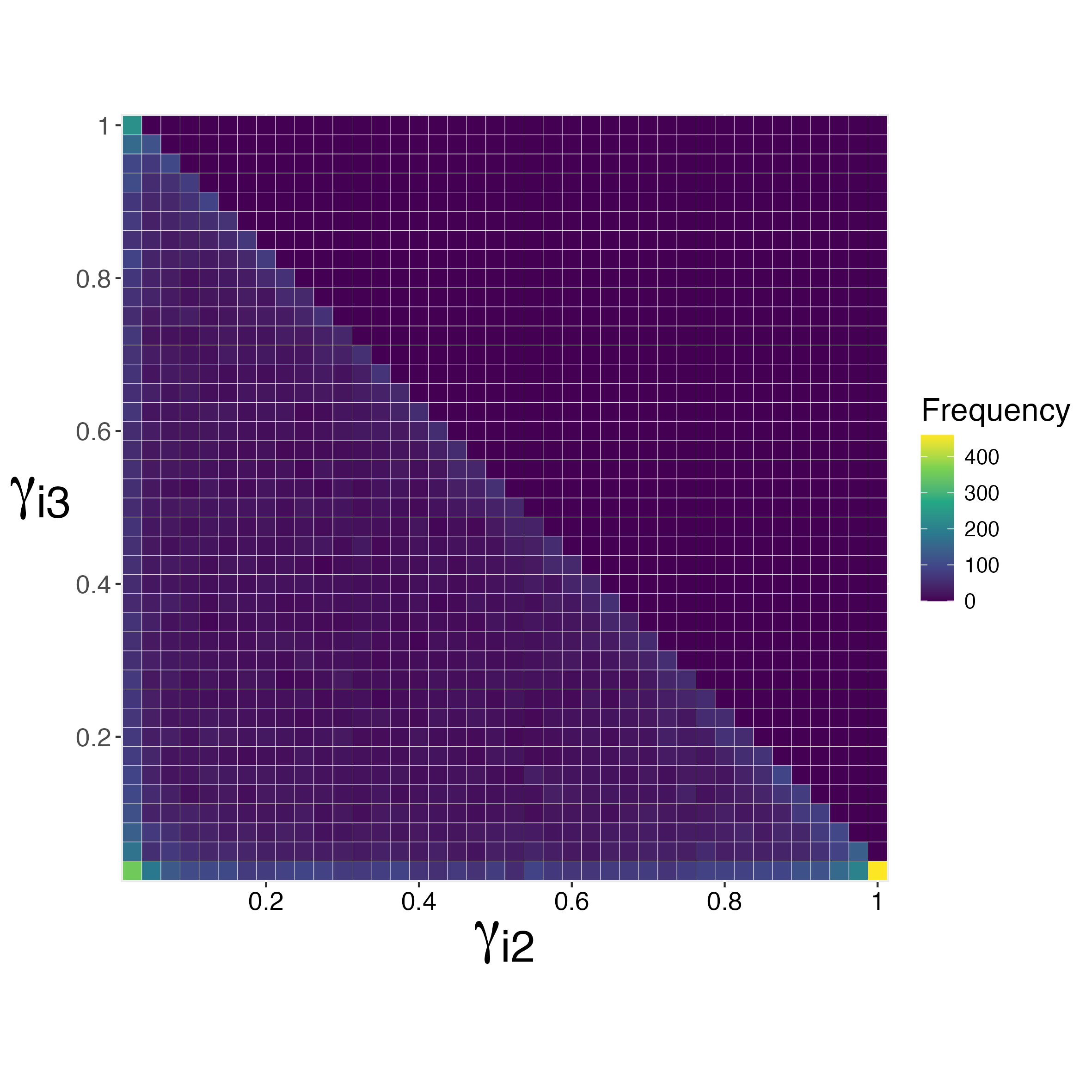}
  \vspace*{-8mm}
  \caption{\centering $\beta = 0.5$}
\end{subfigure}%
\begin{subfigure}{.33\textwidth}
  \centering    
  \vspace*{5mm}
  \includegraphics[trim={0 0 0 2cm},clip,width=1\linewidth]{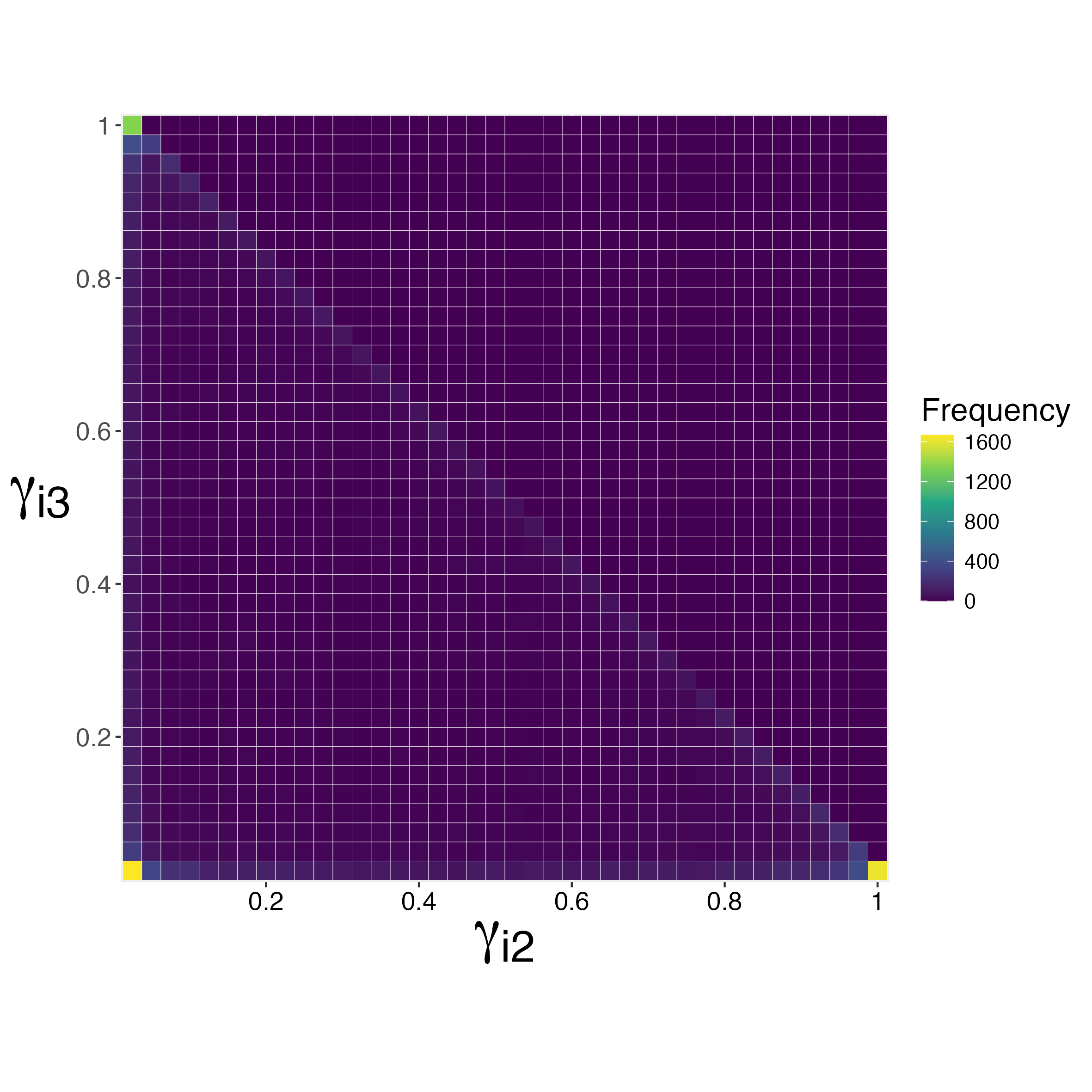}
  \vspace*{-8mm}
  \caption{\centering $\beta = 0.25$}
\end{subfigure}
    \caption{Induced joint distribution of $(\gamma_{i2},\gamma_{i3})$ when the prior distribution of working parameters $\alpha_{0}^{(ij)}$ are tempered at different inverse temperature levels when number of hidden states is $N=3$.}
    \label{fig:induced_transition_probs_tempered_priors}
\end{figure}

Direct application of the general PT algorithm for Bayesian inference of HMMs can lead to computational difficulties when tempering the prior distributions for the working parameters $\alpha_{0}^{(ij)}, \alpha_{1}^{(ij)}$ associated with the transition probabilities. For the normal priors and Gumbel priors introduced in Section \ref{subsec:bayes_hmm} for $\alpha_{0}^{(ij)}$, tempering shifts probability mass in the induced distribution of the transition probabilities $\gamma_{ij}$ toward transition probability matrices with diagonal entries close to one. In other words, it favors configurations near the identity matrix, which can produce near non-ergodic transition structures. This may lead to nearly non-identifiability, which makes the model very difficult to identify. Figure \ref{fig:induced_transition_probs_tempered_priors} illustrates how the probability mass of the induced distribution in the transition probabilities concentrate in certain regions when tempering the priors for the working parameters $\alpha_{0}^{(ij)}$. Moreover, when these regions have non-negligible probability mass in hotter tempered replicas, the high-density regions of the tempered replicas may not overlap, inducing broken ergodicity \citep{stein_broken_1995} and rendering the swap acceptance rate unreliable as a metric for the PT algorithm performance. To address this issue, power-posteriors can be used as tempered replicas instead of power-tempered versions of $\pi$. Namely, if the distribution of interest $\pi(\boldsymbol{x})$ has the form
\begin{equation}
\pi(\boldsymbol{x}|\boldsymbol{y}) \propto p(\boldsymbol{y}\mid\boldsymbol{x}) p(\boldsymbol{x}),
\end{equation}
where $p(\boldsymbol{y} \mid \boldsymbol{x})$ is the likelihood, $\boldsymbol{y}$ are the observations, and and $p(\boldsymbol{x})$ is the joint prior distribution, the power-posteriors are constructed by tempering the likelihood component only \citep{brooks_handbook_2011}:
\begin{equation}
\pi_{\beta}(\boldsymbol{x}|\boldsymbol{y}) \propto [p(\boldsymbol{y}\mid \boldsymbol{x})]^{\beta} p(\boldsymbol{x}).
\end{equation}
See Algorithm \ref{algthm:pt} for the implementation of the reversible PT algorithm when using power-posterior as tempered replicas. For the computation of the power-posteriors for HMMs, see Appendix \ref{appdx:logLik_powerPosterior_algorithms}. 

\begin{algorithm}[ht]
\caption{PT(observations $\left(\boldsymbol{y^{(w)}_{1:T_w}}\right)_w$, latent process $\left(z^{(w)}_{1:T_w}\right)_w$, covariate indicator $c$, inverse temperature schedule $\beta_{0:M}$, initialising chain values $\boldsymbol{X^0} = \left(\boldsymbol{x^{(0)}_{m}}\right)$, number of iterations $H$)}
\begin{algorithmic}[1]
\Require A within-temperature proposal mechanism for each tempered replica $q_m (\boldsymbol{x_m}, \cdot)$.

\vspace{0.2cm}
  \For{$h = 1,\ldots, H$}
    \For{$m = 0,\ldots M$}
        \State Update: $\boldsymbol{x^{(h)}_m} \sim q_m (\boldsymbol{x^{(h-1)}_{m}}, \cdot)$. \Comment{Within-temperature moves} 
    \EndFor

    \State $\boldsymbol{X^{h}} \gets \{\boldsymbol{x^{(h)}_{0}} , \ldots, \boldsymbol{x^{(h)}_{M}} \}$
    \vspace{0.2cm}
    
    \State $k \sim U\{0,\ldots,M-1\}$ 
    \Comment Choose uniformly a pairs of adjacent inverse temperatures
        \vspace{0.2cm}

\Statex \textbf{Compute tempered replicas}
\Comment For definition of LogPowerPosterior($\cdot$), see Appendix \ref{appdx:logLik_powerPosterior_algorithms}
        \vspace{0.2cm}

      \State $\ell_{k} \gets \text{LogPowerPosterior}\left(\left(\boldsymbol{y^{(w)}_{1:T_w}}\right)_w,\left(z^{(w)}_{1:T_w}\right)_w, \boldsymbol{x^{(h)}_k}, c, \beta_k \right)$
      \State $\ell_{k+1} \gets \text{LogPowerPosterior}\left(\left(\boldsymbol{y^{(w)}_{1:T_w}}\right)_w,\left(z^{(w)}_{1:T_w}\right)_w, \boldsymbol{x^{(h)}_{k+1}}, c, \beta_{k+1} \right)$
      \State $\hat{\ell}_k \gets \text{LogPowerPosterior}\left(\left(\boldsymbol{y^{(w)}_{1:T_w}}\right)_w,\left(z^{(w)}_{1:T_w}\right)_w, \boldsymbol{x^{(h)}_{k+1}}, c, \beta_k \right)$
      \State $\hat{\ell}_{k+1} \gets \text{LogPowerPosterior}\left(\left(\boldsymbol{y^{(w)}_{1:T_w}}\right)_w,\left(z^{(w)}_{1:T_w}\right)_w, \boldsymbol{x^{(h)}_k}, c, \beta_{k+1} \right)$

        \vspace{0.2cm}

      \State $a_{k} \leftarrow \min\!\left(0,\; (\tilde{\ell}_i + \tilde{\ell}_j)- (\ell_i + \ell_j)\right)$
      \Comment{Metropolis ratio to accept/reject swap proposal}
      
        \vspace{0.2cm}
      \State $U \sim \text{Unif}(0,1)$
      \If{$\log(U) < a_{k}$}
      
        %\State swap $x^{(k)} \leftrightarrow x^{(k+1)}$
        \State $\boldsymbol{X^{h}} = 
        \{\boldsymbol{x^{(h)}_{0}} , \ldots, \boldsymbol{x^{(h)}_{k+1}}, \boldsymbol{x^{(h)}_{k}},\ldots,  \boldsymbol{x^{(h)}_{M}} \}$
        \Comment Swap chain states if swap accepted
      \EndIf

  \EndFor
    \vspace{0.2cm}

  \State \Return $\{\boldsymbol{X^0},\boldsymbol{X^1},\ldots,\boldsymbol{X^H}\}$.
\end{algorithmic}
\label{algthm:pt}
\end{algorithm}

For the within-temperature steps, we use the component-wise Metropolis–Hastings (CWMH) algorithm. The CWMH algorithm is a variant of Metropolis–Hastings that updates one sub-block of parameters at a time. The dimension of the parameter space in HMMs grows quickly as additional model features are incorporated or as the number of hidden states increases, making the CWHMH algorithm ideal for the within-temperature move steps. See Appendix \ref{appdx:cwmh} for the implementation of the CWMH algorithm for HMMs.

For the inverse temperature schedule, the inverse temperatures are selected to achieve a swap acceptance rate of 0.234 between adjacent inverse temperatures, as discussed in Section \ref{subsec:pt}. The hottest replica $\pi_{\beta_M}$ for the application of the PT algorithm to HMMs is proposed to be a power-posterior with $\beta_M > 0$. Specifically, since we require $\beta_M$ to be small enough for the corresponding hottest tempered replica to be unimodal, or at least easily sampled using standard MCMC routines, the choice of $\beta_M$ was based on candidate hottest replicas that satisfy one of these conditions. To identify an appropriate hottest inverse temperature value, we ran standard MCMC routines to explore the joint posterior distribution of the corresponding power-posterior for each candidate. We selected $\beta_M$ such that the resulting high-density regions of the marginal densities were sufficiently close, sometimes becoming effectively unimodal, allowing for consistent and frequent transitions between them.

\subsubsection*{Practical considerations when setting up PT}

A proper scaling of the proposal distribution $q_{\beta_m} ( \cdot \mid \boldsymbol{x_m})$ for the chains targeting the tempered replicas $\pi_{\beta_m}$ plays a crutial role. There is a vast literature on how to choose an appropriate scaling for different scenarios and proposal schemes; see, for example, \citep{brooks_handbook_2011}, Chapter 4. Although the swap acceptance rates are, in principle, not directly affected by poor mixing of the Markov chains targeting the tempered replicas, and thus the inverse temperature schedule can be constructed independently of the tuning of $q_{\beta_m}(\cdot \mid \boldsymbol{x}_m)$, inadequate within-temperature mixing can substantially increase the number of iterations required to explore each tempered replica effectively. This has a direct impact on the running weight estimates, leading to higher computational costs for accurately estimating the weight of each mode.

The choice of the hottest inverse temperature $\beta_M$ is another important factor. The smaller its value, the more likely $\pi_{\beta_M}$ can be explored successfully via standard MCMC routines, particularly when considering power-posteriors of $\pi$. If the joint prior distribution accounts for the potential within-correlation present in the observed process, a natural candidate for the hottest inverse temperature is $\beta_N = 0$. This choice can simplify the selection of the hottest inverse temperature, as it corresponds to using the joint prior distribution as the hottest replica. However, when this is not the case in the context of HMMs, setting $\beta_M = 0$ will likely lead to negligible swap acceptance rates between $\beta_M$ and any reasonable choice of $\beta_{M-1}$.

Additionally, the smaller $\beta_M$, the more intermediate inverse temperatures between $\beta_0$ and $\beta_M$ may be required to preserve a uniform swap acceptance rate. \citep{nadler_generalized_2007} showed that, under regularity conditions and for a large number of inverse temperatures, the exchange of information between replicas through swap proposals exhibits behaviour analogous to a simple random walk. In that scenario, the expected number of accepted swaps required to complete a single round trip is of order $O(M^2)$ \citep{diaconis_analysis_2000} as $M$ goes to infinity. Furthermore, \citep{syed_non-reversible_2022} showed that, under the DEO scheme, this reduces to $O(M)$. Nonetheless, aiming for large $M$ is not optimal. \citep{roberts_quantifying_2025} showed that, under strong assumptions and in the limit as the dimension $d\rightarrow \infty$ of the state space $\mathcal{X}$, the maximum efficiency of the non-reversible PT algorithm is approximately 42\% higher than that of the reversible PT algorithm.

Selecting a hottest replica $ \pi_{\beta_M}$ for which there remains a substantial gap between high-density regions can also significantly increase the computational cost of estimating the weights of each mode using running weight estimates. We suggest the use of analytical results together with exploratory Markov chain routines targeting candidate $\pi_{\beta_M}$ to examine the geometry of the distribution before finalizing the choice of the hottest replica.

\section{Application to ecological time series}\label{sec:results}

We use the implementation guidelines developed in Section \ref{subsec:bayes_hmm} to implement the reversible PT algorithm and analyze blue whale dive data using HMMs. The dataset was previously analyzed within an HMM framework in \citep{deruiter_multivariate_2017} to quantify the impact of sound stimuli on blue whale behaviour. Inference was conducted using maximum likelihood estimation, and multiple local maxima were reported during the model fitting procedure, making this a suitable setting to test and compare our approach. We propose two Bayesian HMMs to analyse the blue whale dive data: a baseline 5-dimensional 3-state HMM with a time-homogeneous transition probability matrix, and an extension that incorporates sound stimuli during dives as a covariate in the transition probability matrix.

\subsection{Bayesian HMMs for blue whale movements}\label{subsec:blue_whale_data}

Animal-borne tags were deployed to track the movements of 37 blue whales, and the collected information was processed into multiple data streams. See Supplement 2 from \citet{deruiter_multivariate_2017} for a full description of the data collection and data processing steps. While the blue whales were tracked, they were also exposed to sound stimuli, including mid-frequency military sonars and pseudo-random noise in the same frequency range in order to understand how the stimuli affects their movements, and thus behaviors. 

Five data streams from the full dataset were considered for analysis: the number of lunges, dive duration, post-dive surface duration, maximum depth, and step length (in the horizontal dimension). For construction of the state-dependent distributions, we assume contemporaneous conditional independence; see \citep{zucchini_hidden_2017} for explicit assumption formulation. Consistent with the model applied in \cite{deruiter_multivariate_2017}, continuous, non-negative data streams were modeled with a Gamma distribution, parameterized in terms of the mean and standard deviation, while data streams with positive integer outcomes were modeled with a Poisson distribution:
\begin{equation} \label{eq:sdd_bluewhale}
\begin{aligned}
    \text{Number of lunges: }y_{t1} &\mid s_t \sim \text{Poisson}(\lambda_{s_t}), & \phantom{s_t = 1, 2, 3}\\
    \text{Dive duration: }y_{t2} &\mid s_t \sim \text{Gamma}(\mu_{2s_t},\sigma_{2s_t}), & \phantom{s_t = 1, 2, 3}\\
    \text{Surface duration: }y_{t3} &\mid s_t \sim \text{Gamma}(\mu_{3s_t},\sigma_{3s_t}), & s_t = 1, 2, 3.\\
    \text{Maximum depth: }y_{t4} &\mid s_t \sim \text{Gamma}(\mu_{4s_t},\sigma_{4s_t}), & \phantom{s_t = 1, 2, 3}\\
    \text{Step length: }y_{t5} &\mid s_t \sim \text{Gamma}(\mu_{5s_t},\sigma_{5s_t}), & \phantom{s_t = 1, 2, 3}
\end{aligned}
\end{equation}
For the baseline 3-state HMM, a time-homogeneous process for $\boldsymbol{\Gamma}$ was assumed. The reparameterization from Equation \ref{eq:tpm_cov} was used for the transition probabilities, and the working parameters $\alpha_{ij}$ were estimated instead of the transition probabilities on their original scale. Weakly informative priors were chosen for the state-dependent distribution parameters; see Appendix \ref{appdx:priors_implementation} for the explicit formulations. For the initial state distribution vector $\boldsymbol{\delta}$ it was assumed to follow a $\text{Dirichlet}(\boldsymbol{1}_N)$ as prior. For the working parameters $\alpha_{0}^{(ij)}$, the priors were specified as introduced in Section \ref{subsec:bayes_hmm}. In total, the baseline 3-state HMM required estimating 36 parameters, plus 3 latent variables related to the prior of $\alpha_{0}^{(ij)}$. 

For the extended baseline model, the exogenous variable indicating the presence of sound stimuli to which whales were exposed during tracking was incorporated as a covariate in the transition probabilities via a multinomial logit transformation, as formulated in equation \ref{eq:tpm_cov}. This extension breaks the time-homogeneity assumption of the hidden process, resulting in a non-homogeneous HMM \citep{zucchini_hidden_2017}. The specifications for the state-dependent distributions remained the same as in the baseline model, and the same priors formulated in the Appendix \ref{appdx:priors_implementation} were used. Since the variable indicating the presence of sound stimuli is binary, we retain the notation from Equation \ref{eq:tpm_cov} in Section \ref{sec:hmm} throughout this section, without introducing auxiliary variables as in Equation \ref{eq:tpm_cov_categorical} in Section \ref{subsec:bayes_hmm}. The priors for the covariate coefficients $\alpha_{1}^{(ij)}$ introduced in Section \ref{subsec:bayes_hmm} were adopted for these parameters. In total, the extended 3-state HMM required estimation of 42 parameters, plus 3 latent variables related to the prior on the working parameters.

\subsection{Parallel tempering algorithm}\label{subsec:pt_applied_hmms}

The inverse-temperature schedules for implementing the PT algorithm for the two proposed Bayesian HMMs were tuned to achieve a uniform swap acceptance rate across tempered replicas, as introduced in Section \ref{subsec:pt}. Specifically, the schedules were tuned to achieve a swap acceptance rate between adjacent temperatures of 0.23. Initially, the Robbins–Monro algorithm \citep{robbins_stochastic_1951} was used to estimate the next hottest inverse temperature, targeting a swap acceptance rate of 0.234. However, this approach proved unstable due to the introduction of sensitive tuning parameters. As a result, the schedule was constructed by adding one inverse temperature at a time. Once the current set achieved the desired swap acceptance rate between adjacent inverse temperatures, the process continued until reaching an adequate hottest temperature. For each proposed new hottest inverse temperature, the PT algorithm was run for 50,000 iterations using the extended schedule, and the swap acceptance rate between the proposed and current hottest temperatures was monitored to determine whether the candidate should be closer to or farther from the current value. The new hottest inverse temperature was selected once this rate was sufficiently close to 0.234, specifically between 0.22 and 0.24.

For the within-temperature move steps, a single update was performed between swap proposals, that is $U=1$. Additionally, the component-wise Metropolis-Hastings algorithm was used to marginally explore the parameter space of the tempered replicas. The proposal distributions for each implementation were tuned to achieve a step size that induced a proposal acceptance rate between 0.2 and 0.4 for each sub-block of the full parameter vector.

The parallel tempering algorithm was implemented in C++ via the Rcpp R package, which provides R functions as well as C++ classes that enable seamless integration of R and C++ \citep{eddelbuettel_rcpp_2011}. Additionally, parallel computing was incorporated via the RcppParallel R package, reducing computational cost per algorithm implementation by a factor of 10. The PT algorithm was implemented on the Digital Research Alliance of Canada (DRAC) servers by submitting jobs in batches of 400,000 iterations for each PT algorithm implementation. For both models, 10 PT algorithm implementations were carried out, each aiming for 2,000,000 iterations, and all PT algorithm implementations per model were initialized using random starting values. Once all batches were completed, the states from the coldest chain were extracted from all PT algorithm implementations and merged in a post-sampling process. 

For both models, the first 1,000,000 iterations from the coldest chain of each PT algorithm implementation were discarded as burn-in, and the remaining 1,000,000 iterations were retained as samples. After removing burn-in, a visual inspection of the marginal posterior distributions revealed label switching in all 10 coldest chains for both models. A further visual inspection of the marginal posterior distributions based on samples from the coldest chain was conducted to identify parameters with well-separated high-density regions that could be used to correct label switching via artificial ordering constraints. We found that the estimated mean maximum depth $\mu_{4n}$ for each hidden state $n$ exhibited clearly separated high-density regions. Based on this, the label switching present in the posterior samples was corrected by reordering the hidden state labels according to an ordering constraint on the state-dependent distribution parameters associated with maximum dive depth for all coldest chains in both models, specifically on the estimated mean maximum depth for each hidden state:
\begin{equation}\label{eq:ordering_constraint}
\mu_{41} < \mu_{42} < \mu_{43}.    
\end{equation}
For both the baseline 3-state HMM and its extension, multiple high-density regions were identified after correcting for label switching, indicating genuine multimodality. For each model, summary statistics for the marginal posterior distributions, such as the posterior median and 95\% credible intervals, were computed mode-wise. That is, for each mode, the summary statistics were calculated using only samples from that mode, without incorporating parameter estimates from other high-density regions. We relied on the auxiliary variables $\hat{w}_{h,K}$, introduced in Section \ref{subsec:genuine_multimodality}, which are used to estimate the weight of a given mode $h$, to filter for samples corresponding to each specific mode.

\subsubsection*{Baseline 3-state HMM}

For the baseline model, a 12-temperature inverse schedule was constructed with the hottest temperature set to $\beta_M = 0.019$. See Table \ref{tab:base_hmm} for the complete 12-temperature schedule. The job wall-clock time of each 400,000-iterations batch per PT algorithm implementation ranged between 5–15 hours, and with the job submission queue wait times, the full number of iterations concluded in four days.

\begin{table}[htp!]
\centering
\begin{tabular}{cccccccccccc}
\hline
$\beta_0$ & $\beta_1$ & $\beta_2$ & $\beta_3$ & $\beta_4$ & $\beta_5$ & $\beta_6$ & $\beta_7$ & $\beta_8$ & $\beta_9$ & $\beta_{10}$ & $\beta_{11}$ \\ %\hline
1 & 0.667 & 0.444 & 0.298 & 0.2 & 0.139 & 0.096 & 0.068 & 0.05 & 0.039 & 0.027 & 0.019 \\ \hline
\end{tabular}
\caption{12-inverse temperature schedule selected for the baseline 3-state HMM.}
\label{tab:base_hmm}
\end{table}

\begin{figure}[htp!]
    \centering
    \includegraphics[scale=.5]{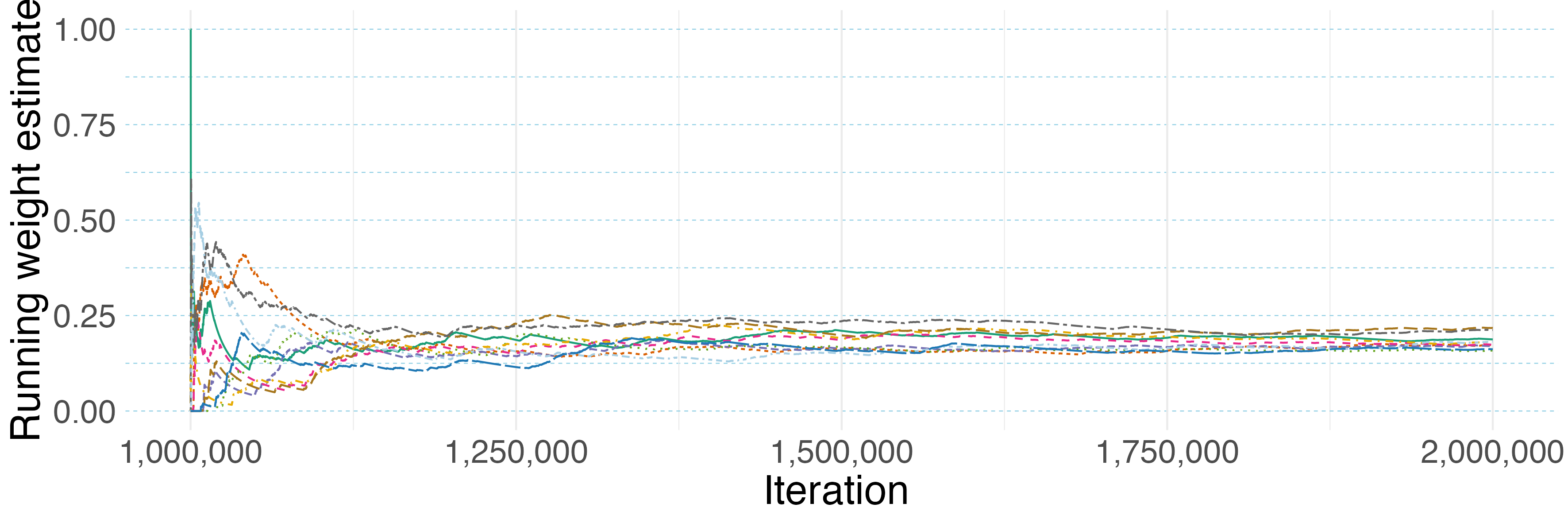}
    \caption{Running weight estimates $\hat{w}_{B,K} $ of mode $B$ accross all 10 PT algorithm implementations.}
    \label{fig:weights_baseline}
\end{figure}

Consistent swap acceptance rates across all runs, along with traceplot of the information from the coldest replica moving across the tempered replicas were indicative of a successful implementation of the algorithm. Additionally, a standard deviation of 4.77 in the total number of round trips across the different PT algorithm implementations suggests that the PT algorithm performed consistently across all runs. See Table \ref{tab:baseline_roundtrip} for details. For a visualization of how information from the coldest replica moves across all tempered replicas for each PT algorithm implementation, see Appendix \ref{appdx:diagnostics}. Additionally, $\hat{R}$ and effective sample sizes were computed for the estimated parameters using samples from the coldest chain, after correcting for label switching in the post-sampling process via the ordering constraint defined in Equation \ref{eq:ordering_constraint}; for these metrics, see Appendix \ref{appdx:rhat_ess}.
\begin{table}[htp!]
    \centering
    \begin{tabular}{rcccccccccc}
    \hline
        PT implementation ID & 1 & 2 & 3 & 4 & 5 & 6 & 7 & 8 & 9 & 10 \\ %\hline
        Total round trips & 90 & 90 & 78 & 80 & 87 & 90 & 89 & 89 & 92 & 91 \\ \hline
    \end{tabular}
    \caption{Total number of round trips per PT algorithm implementation corresponding to the baseline 3-state HMM.}
    \label{tab:baseline_roundtrip}
\end{table}
After correcting for label switching across all chains, two high-density regions, denoted as mode $A$ and $B$, were identified in the coldest chain states. Running weight estimates were used for the estimation of the mode weight associated to the high-density region $B$. The running weight estimates for mode $B$, $\hat{w}_{B,K}$, were computed using the estimated values of $\mu_{43}$, which corresponds to the mean paramater for the state-depedent distribution associated with the maximum depth per dive for hidden state $n = 3$. Table \ref{tab:baseline_weigth_modeA} shows the running weight estimates for mode $B$ across all PT algorithm implementations and Figure \ref{fig:weights_baseline} illustrates their convergence. The average of these estimates is 0.181 with a standard deviation of 0.01976.  

\begin{table}[htp!]
    \centering
    \scalebox{.94}{
    \begin{tabular}{rcccccccccc}
    \hline
        {\footnotesize PT  implementation ID}  & 1 & 2 & 3 & 4 & 5 & 6 & 7 & 8 & 9 & 10 \\ %\hline
        $\hat{w}_{B,K} $ & 0.187 & 0.172 & 0.170 & 0.174 & 0.158 & 0.180 & 0.218 & 0.212 & 0.182 & 0.162 \\ \hline
    \end{tabular}
    }
    \caption{Running weight estimates $\hat{w}_{B,K} $ of mode $B$.}
    \label{tab:baseline_weigth_modeA}
\end{table}

\begin{figure}[htp!]
%\vspace*{-5mm}
%\vspace*{-7mm}
\begin{subfigure}{.33\textwidth}
  \centering
    \includegraphics[trim={0cm 0 0 0cm},clip,width=1\linewidth]{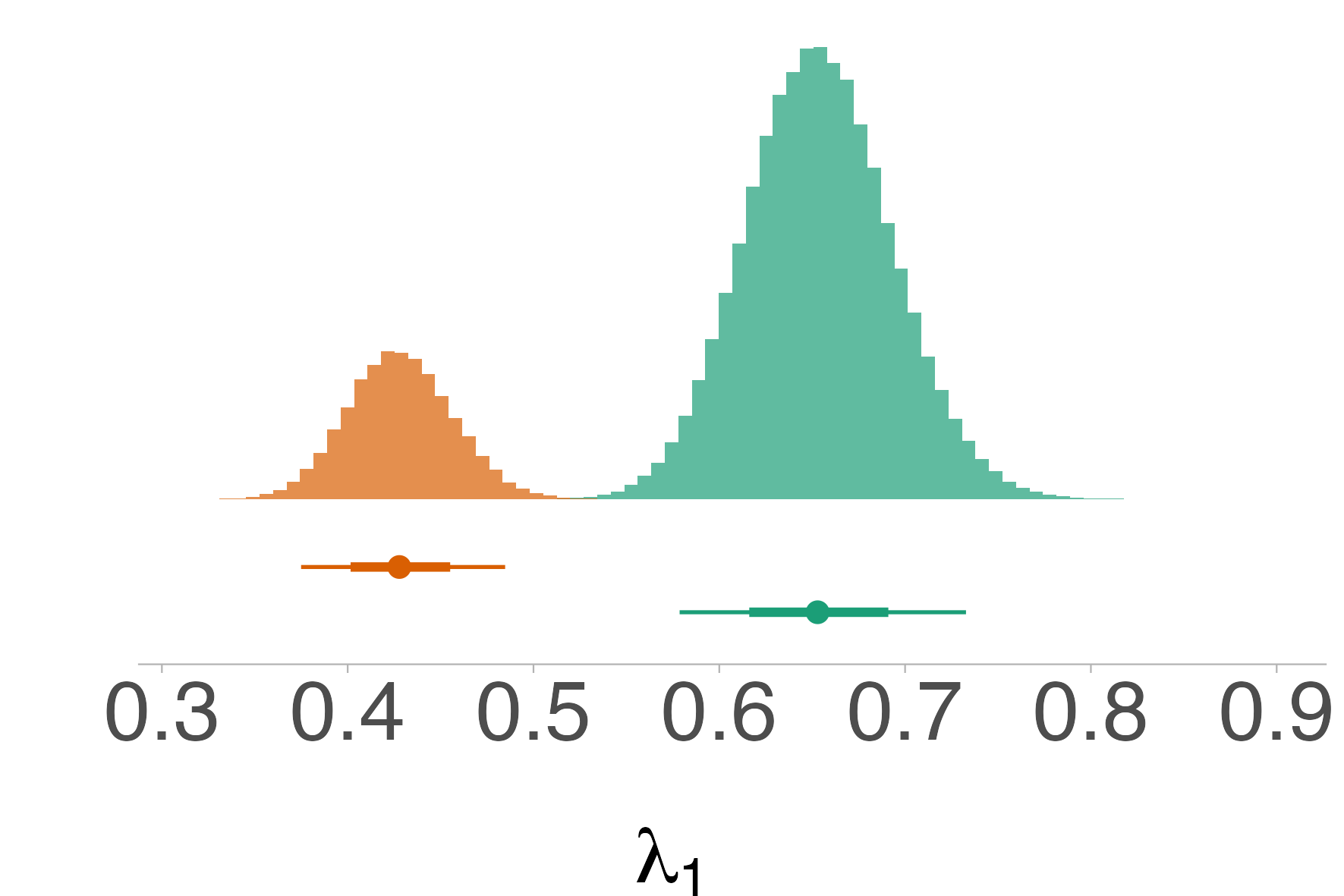}
\end{subfigure}%
\begin{subfigure}{.33\textwidth}
  \centering
    a) \includegraphics[trim={0 0 0 0cm},clip,width=1\linewidth]{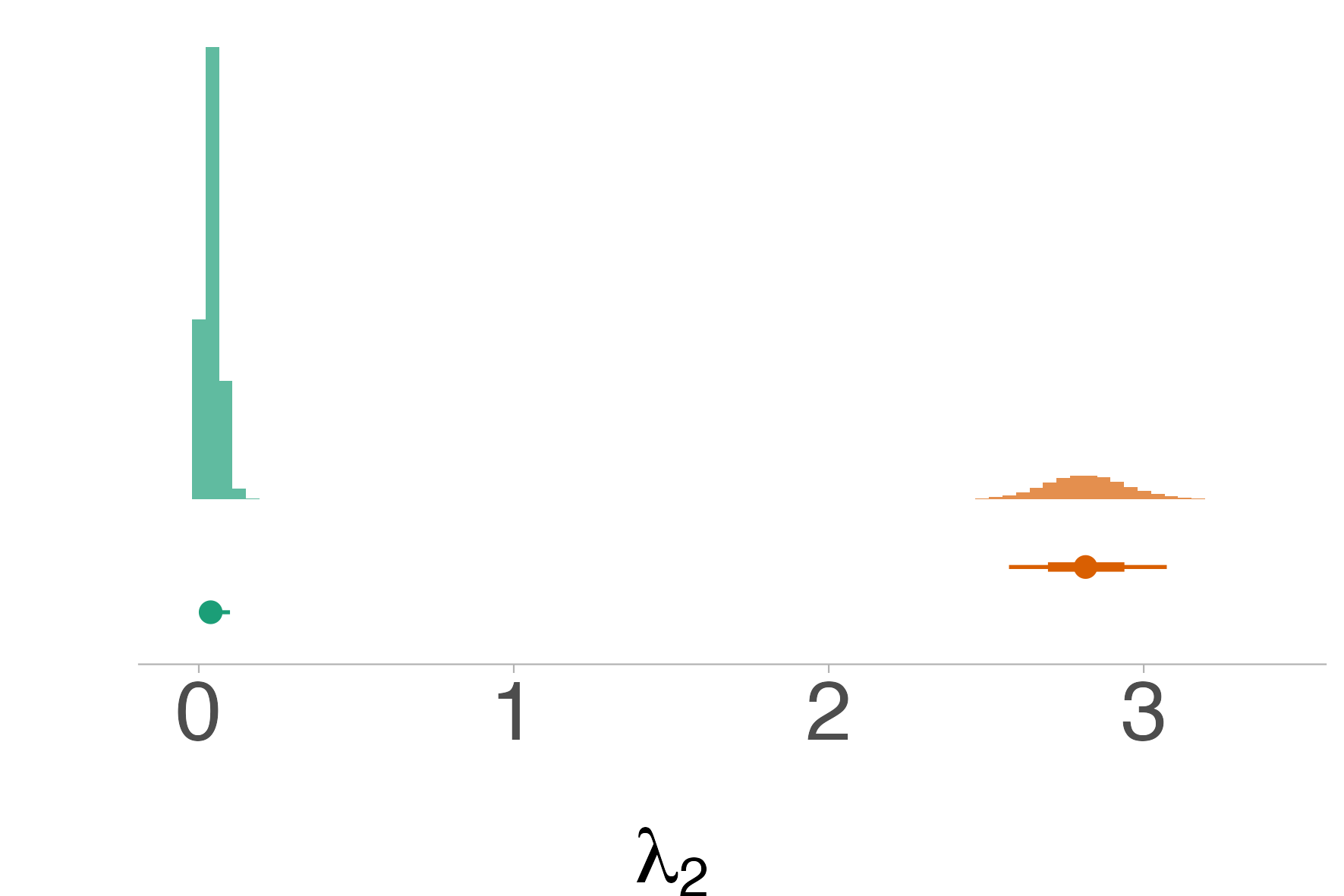}
\end{subfigure}
\begin{subfigure}{.33\textwidth}
  \centering
     \includegraphics[trim={0 0 0 0cm},clip,width=1\linewidth]{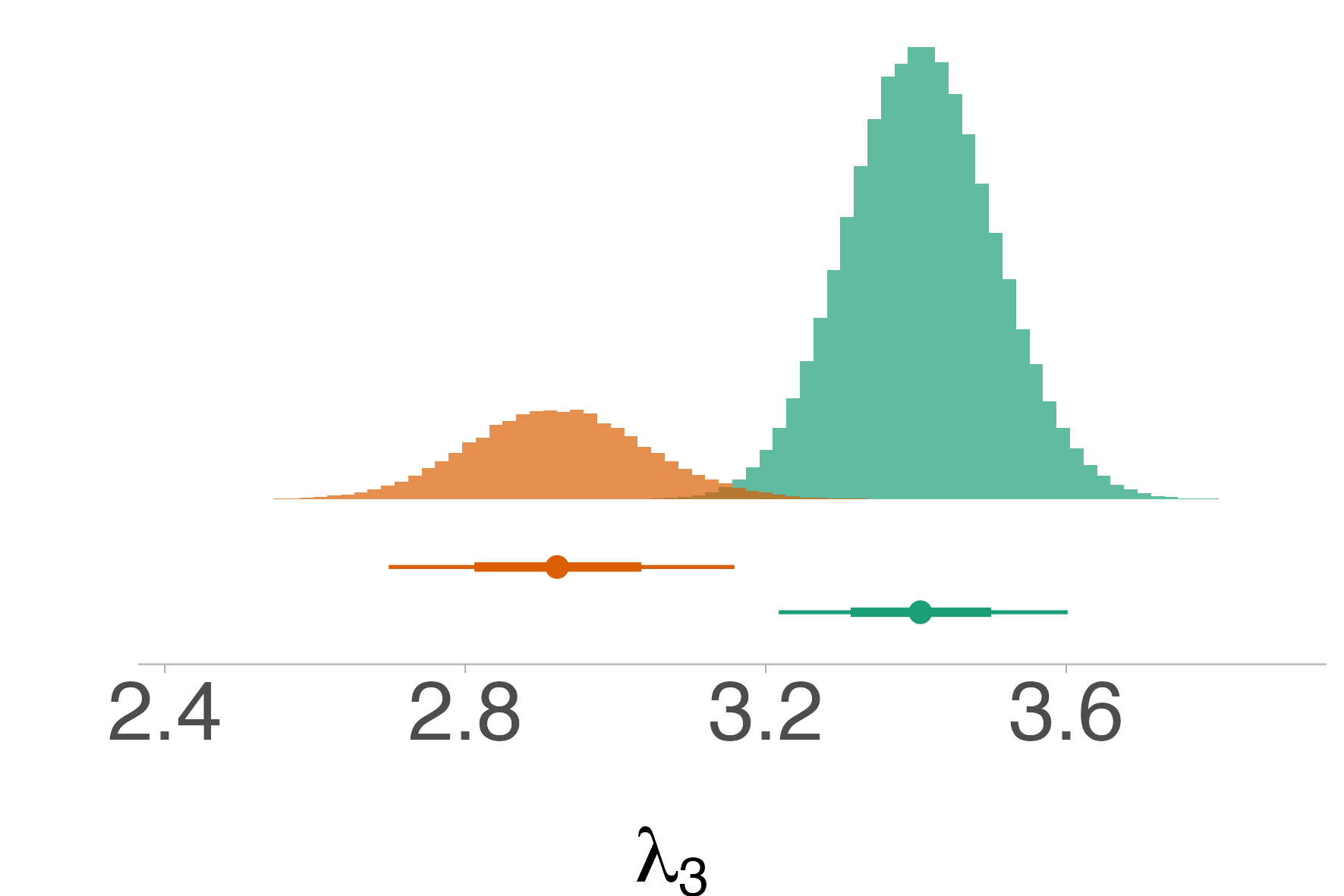}
\end{subfigure}

\vspace*{5mm}
\begin{subfigure}{1.0\textwidth}
  \centering
  b)
  \includegraphics[trim={0 0 0 0cm},clip,width=1\linewidth]{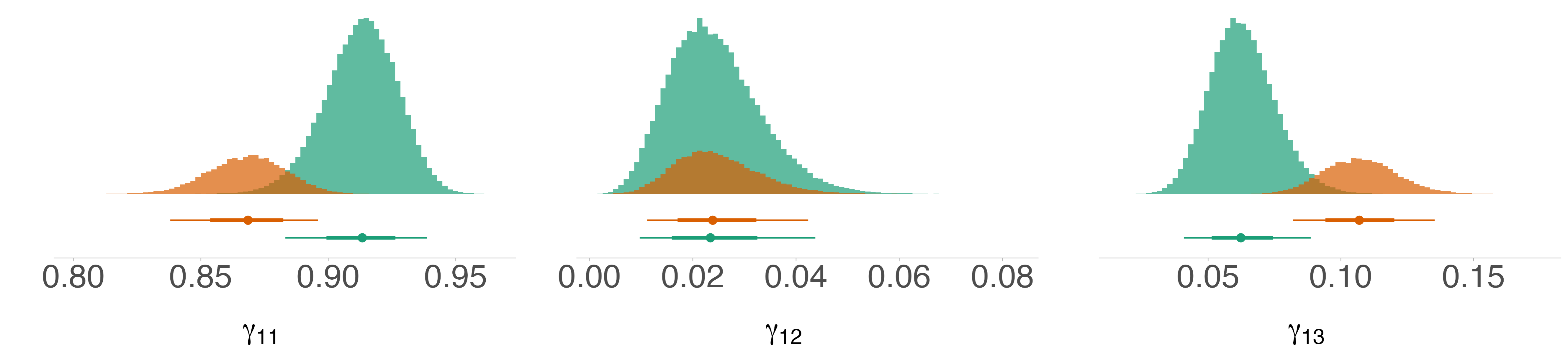}
\end{subfigure}

\vspace*{5mm}
\begin{subfigure}{1.0\textwidth}
  \centering
  c)
  \includegraphics[trim={0 0 0 0cm},clip,width=1\linewidth]{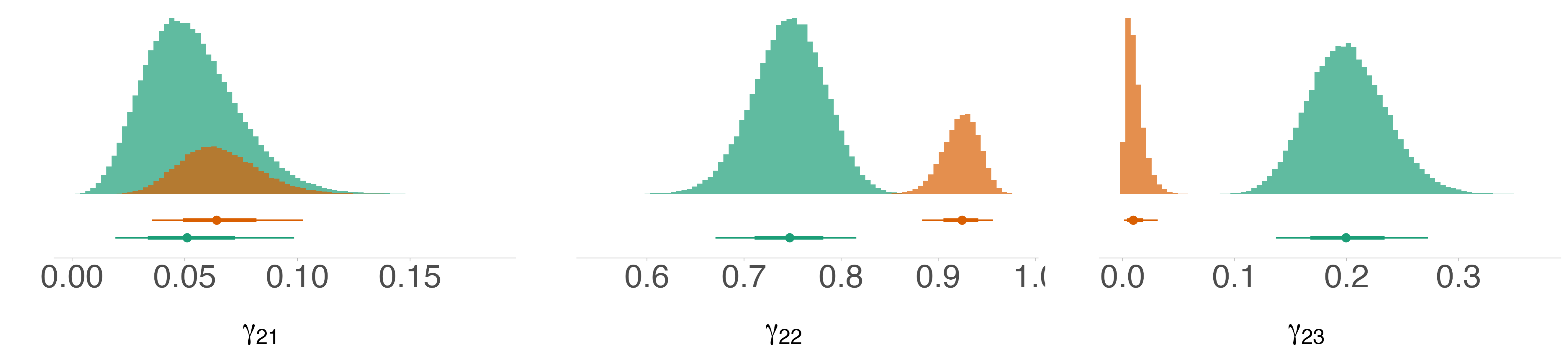}
\end{subfigure}

\vspace*{5mm}
\begin{subfigure}{1.0\textwidth}
  \centering
  d)
  \includegraphics[trim={0 0 0 0cm},clip,width=1\linewidth]{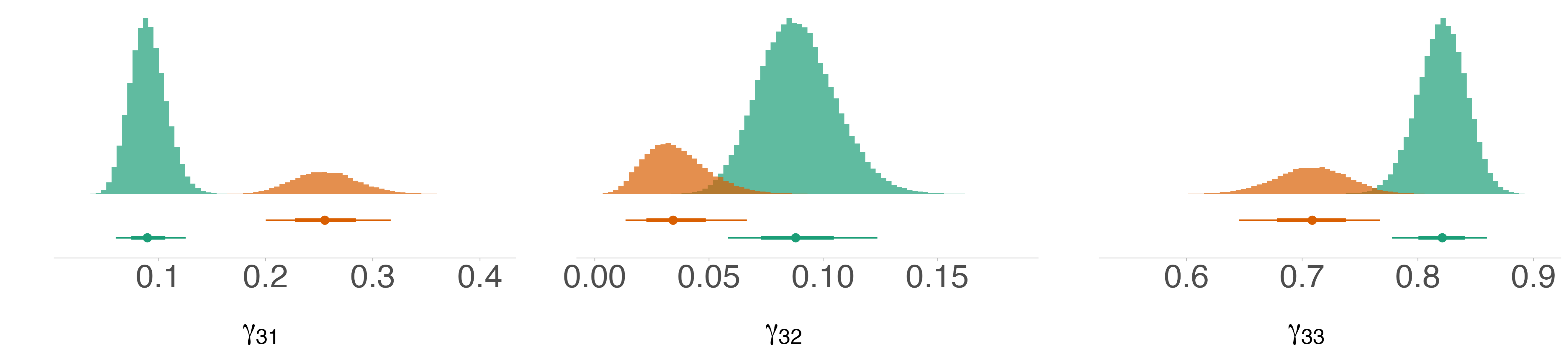}
\end{subfigure}
\caption{Histograms of the marginal posterior distributions estimated from the coldest chain. Row a) corresponds to the rate parameter $\lambda_n$ associated with the number of lunges data stream for each hidden state $n$, while rows b)—d) correspond to the transition probabilities $\gamma_{ij}$. The color indicates the mode to which the high-density region correspond. The lines below the histograms are the mode-wise 95\% credible interval. The dot in the line indicates the posterior median, whereas the thicker line inside the 95\% credible interval indicates the mode-wise 66\% credible interval.}
\label{fig:sddp_lunges}
\end{figure}

The chain states from the coldest chain corresponding to the PT algorithm implementation with ID 1 were extracted for the computation of the marginal posterior distributions, as well as the 66\% and 95\% credible intervals. Figure \ref{fig:sddp_lunges} illustrates the estimated marginal posterior distribution of the state-dependent distribution parameters associated with the number of lunges and the entries of the transition probability matrix.

\subsubsection*{3-state HMM with sound stimuli covariate}

The temperature schedule used to implement the parallel tempering algorithm for this extension resulted in a 13-temperature inverse schedule, with the hottest temperature set to $\beta_M = 0.015$. See Table \ref{tab:base_hmm_cov} for the complete 13-temperature schedule. The job wall-clock time of each 400,000-iterations batch per PT algorithm implementation took about 8–20 hours, and with the job submission queue wait times, the full number of iterations concluded in six days.

\begin{table}[htp!]
    \centering
    \scalebox{0.95}{
        \begin{tabular}{ccccccccccccc}
        \hline
        $\beta_0$ & $\beta_1$ & $\beta_2$ & $\beta_3$ & $\beta_4$ & $\beta_5$ & $\beta_6$ & $\beta_7$ & $\beta_8$ & $\beta_9$ & $\beta_{10}$ & $\beta_{11}$ & $\beta_{12}$ \\ %\hline
        1 & 0.690 & 0.476 & 0.328 & 0.227 & 0.156 & 0.110 & 0.078 & 0.057 & 0.043 & 0.031 & 0.022 & 0.015 \\ \hline
        \end{tabular}
        }
\caption{$13-$inverse temperature schedule selected for the extended baseline 3-state HMM.}
\label{tab:base_hmm_cov}
\end{table}

Consistent swap acceptance rates across runs, along with traceplots showing the movement of information from the coldest replica across the tempered replicas, indicated a successful implementation of the PT algorithm. Additionally, a standard deviation of 6.51 in the total number of round trips across the different PT algorithm implementations suggests that the PT algorithm performed consistently across all runs. Details are provided in Table \ref{tab:baseline_roundtrip_cov}. Traceplots showing how information from the coldest replica moved across all tempered replicas for each PT algorithm implementation are included in Appendix \ref{appdx:diagnostics}. Additionally, $\hat{R}$ and effective sample sizes were computed for the estimated parameters using samples from the coldest chain, after correcting for label switching in the post-sampling process via the ordering constraint defined in Equation \ref{eq:ordering_constraint}; for these metrics, see Appendix \ref{appdx:rhat_ess}.

\begin{table}[htp!]
    \centering
    \begin{tabular}{rcccccccccc}
    \hline
        PT implementation ID & 1 & 2 & 3 & 4 & 5 & 6 & 7 & 8 & 9 & 10 \\ %\hline
        Total round trips & 75 & 53 & 60 & 70 & 57 & 60 & 60 & 66 & 63 & 67 \\ \hline
    \end{tabular}
    \caption{Total number of round trips per PT algorithm implementation corresponding to the extended 3-state HMM with covariates in the transition probabilities.}
    \label{tab:baseline_roundtrip_cov}
\end{table}

    \begin{table}[htp!]
        \centering
\scalebox{.84}{
        \begin{tabular}{rcccccccccc}
        \hline
            {\small PT implementation ID} & 1 & 2 & 3 & 4 & 5 & 6 & 7 & 8 & 9 & 10 \\ %\hline
            $\hat{w}_{\tilde{B},K} $ & 0.0037& 0.0041& 0.0041& 0.0027& 0.0048& 0.0043& 0.0035& 0.0044& 0.0034& 0.0035 \\ \hline
        \end{tabular}
}
        \caption{Running weight estimates $\hat{w}_{\tilde{B},K} $ for mode $\tilde{B}$.}
        \label{tab:baseline_weigth_modeA_cov}
    \end{table}

\begin{figure}[htp!]
    \centering
    \includegraphics[scale=.5]{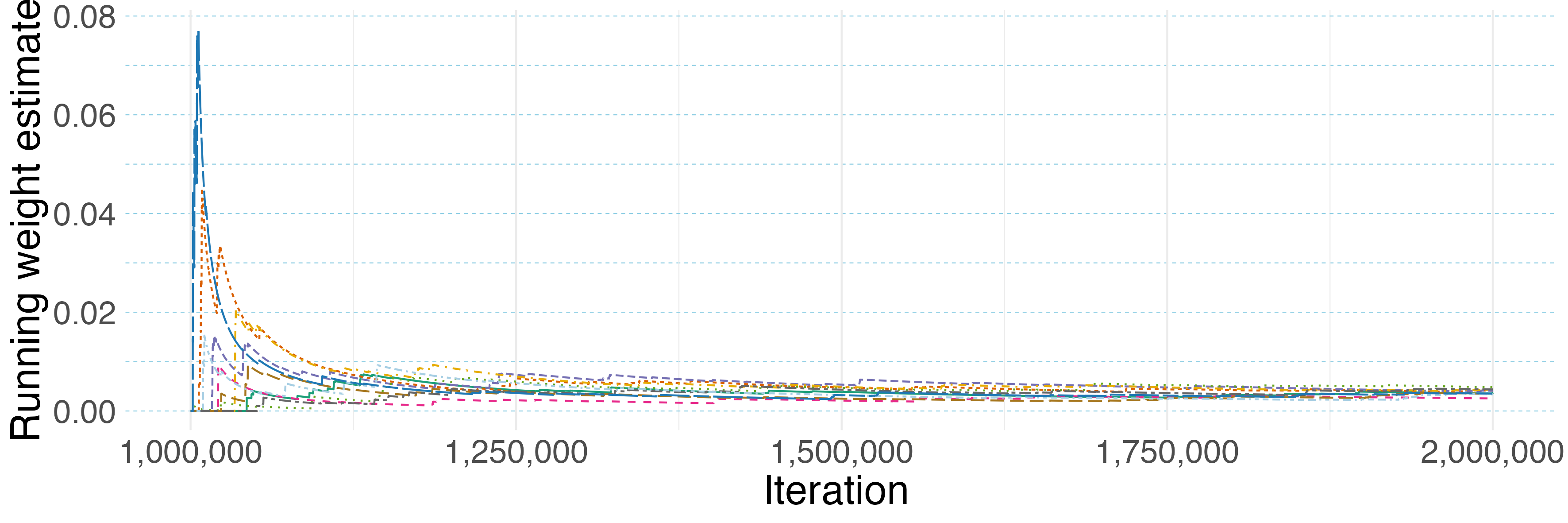}
    \caption{Running weight estimates of mode $\tilde{B}$ across all 10 PT algorithm implementations.}
    \label{fig:weights_baseline_cov}
\end{figure}

After correcting for label switching using the  ordering constraints defined in Equation \ref{eq:ordering_constraint}, two high-density regions, denoted as mode $\tilde{A}$ and $\tilde{B}$, were identified in the coldest chain states. Running weight estimates were used for the estimation of the mode weight associated to the high-density region $\tilde{B}$. The running weight estimates for mode $\tilde{B}$, $\hat{w}_{\tilde{B},K}$, were computed using the estimated values of $\mu_{43}$, which corresponds to the mean parameter for the state-dependent distribution associated with the maximum depth per dive for hidden state $n = 3$. Table \ref{tab:baseline_weigth_modeA_cov} shows the running weight estimates for mode $\tilde{B}$ across all PT algorithm implementations and Figure \ref{fig:weights_baseline_cov} illustrates their convergence. The average of these estimates is 0.00385 with a standard deviation of 0.00061.

\begin{figure}[htp!]
%\vspace*{-5mm}
%\vspace*{-7mm}
\begin{subfigure}{.33\textwidth}
  \centering
    \includegraphics[trim={0cm 0 0 0cm},clip,width=1\linewidth]{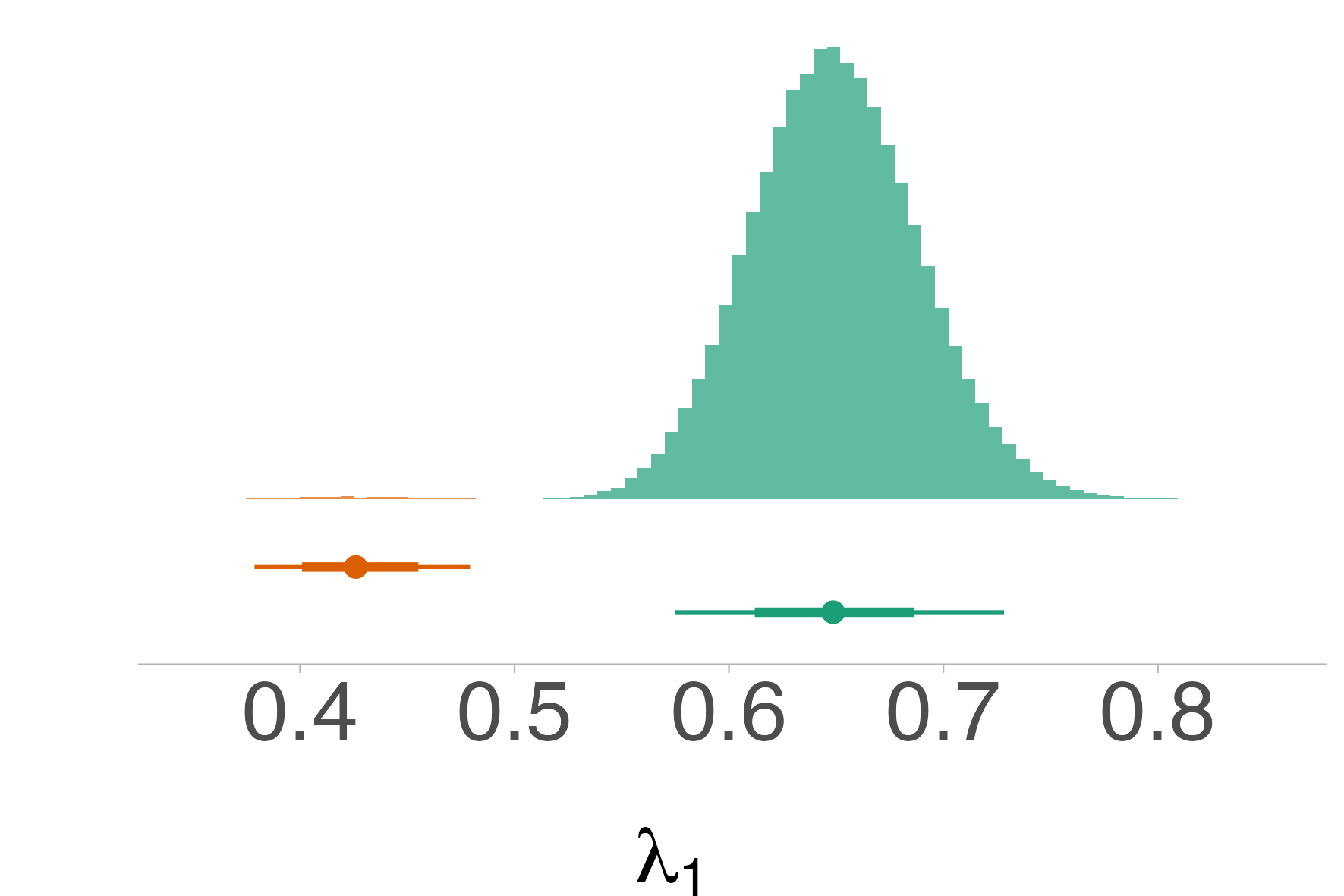}
\end{subfigure}%
\begin{subfigure}{.33\textwidth}
  \centering
    a) \includegraphics[trim={0 0 0 0cm},clip,width=1\linewidth]{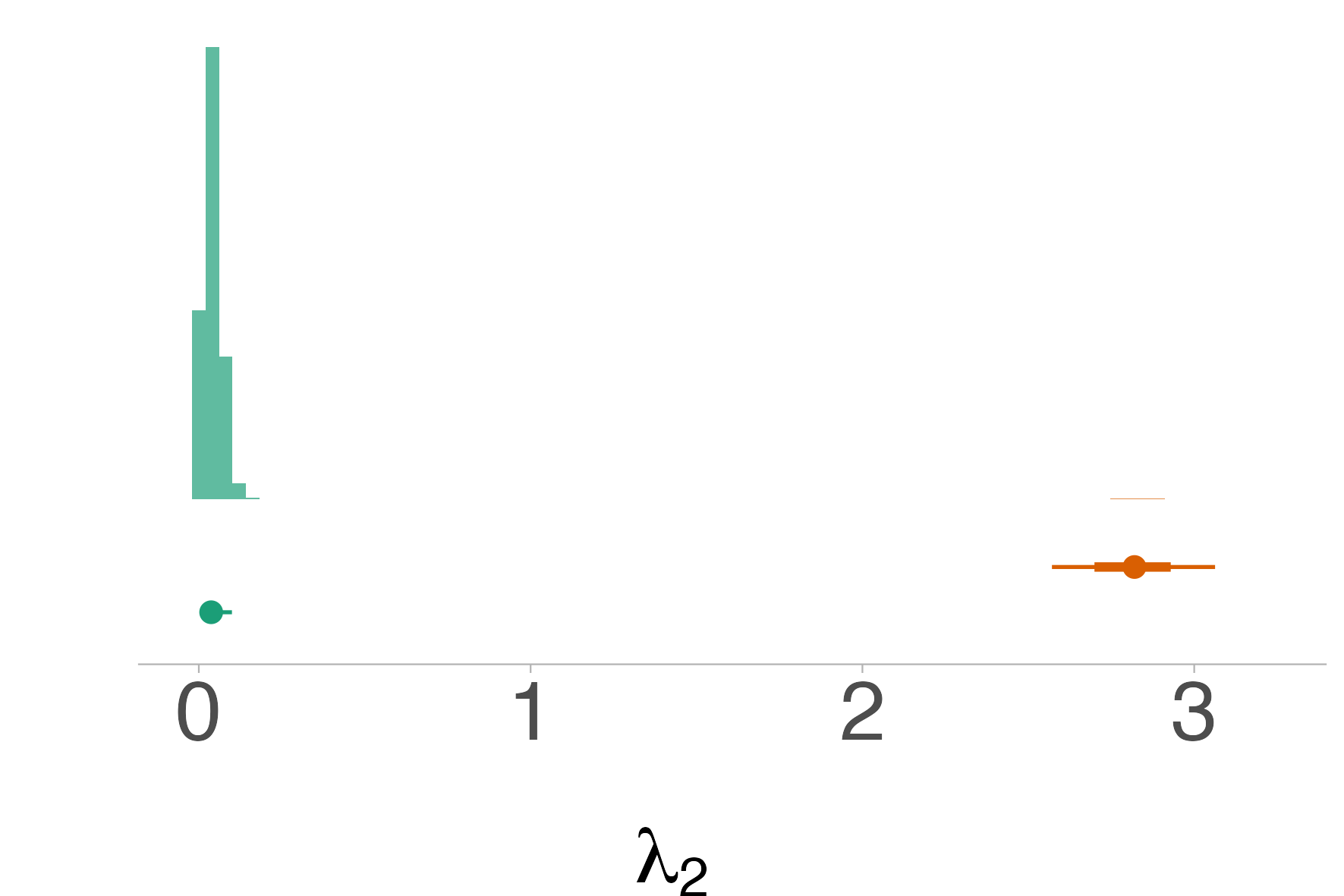}
\end{subfigure}
\begin{subfigure}{.33\textwidth}
  \centering
     \includegraphics[trim={0 0 0 0cm},clip,width=1\linewidth]{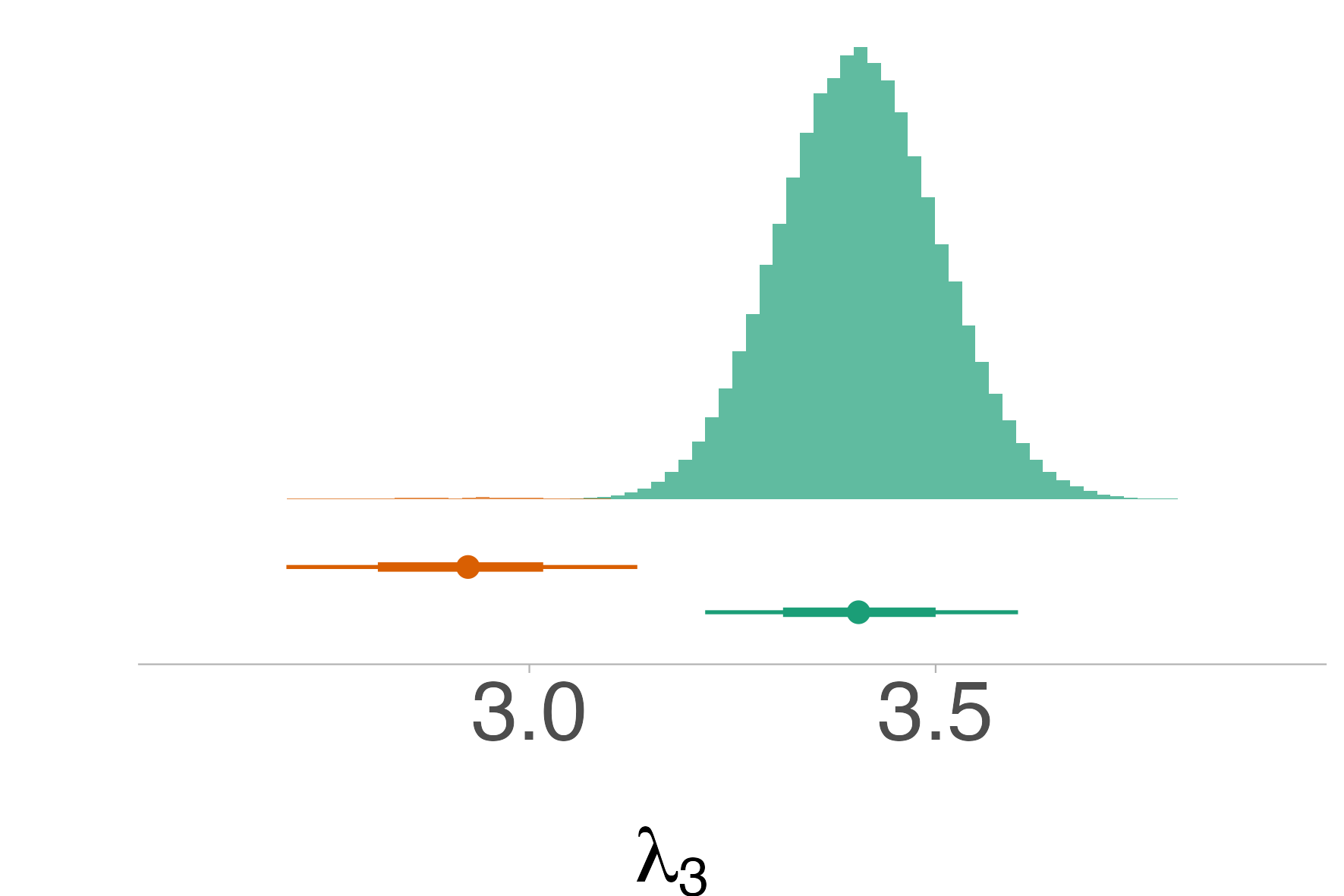}
\end{subfigure}

\vspace*{5mm}
\begin{subfigure}{1.0\textwidth}
  \centering
  b)
  \includegraphics[trim={0 0 0 0cm},clip,width=1\linewidth]{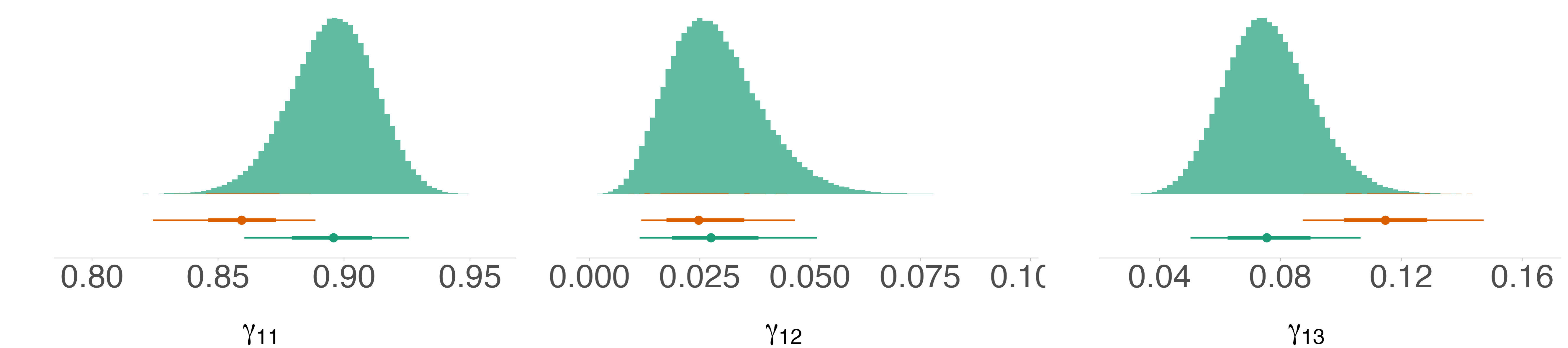}
\end{subfigure}

\vspace*{5mm}
\begin{subfigure}{1.0\textwidth}
  \centering
  c)
  \includegraphics[trim={0 0 0 0cm},clip,width=1\linewidth]{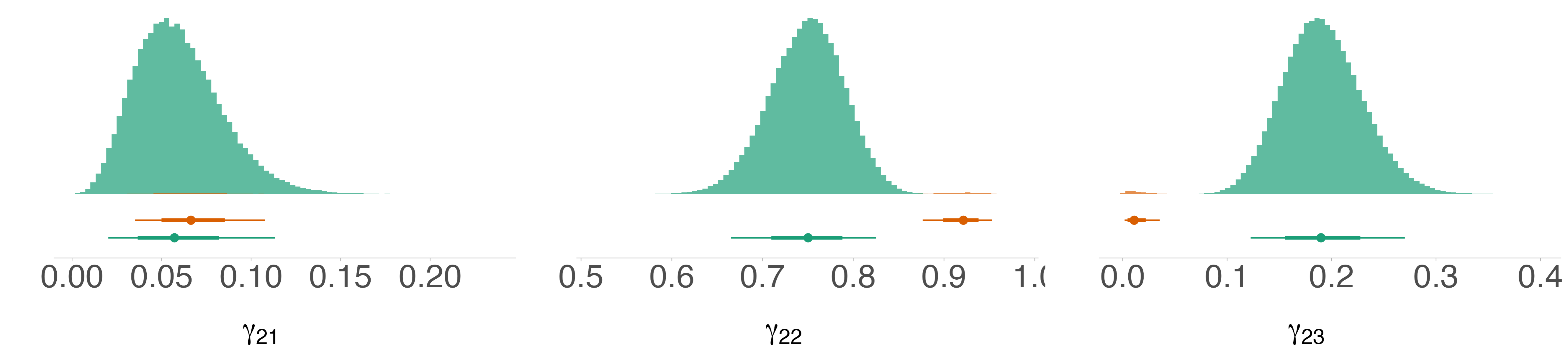}
\end{subfigure}

\vspace*{5mm}
\begin{subfigure}{1.0\textwidth}
  \centering
  d)
  \includegraphics[trim={0 0 0 0cm},clip,width=1\linewidth]{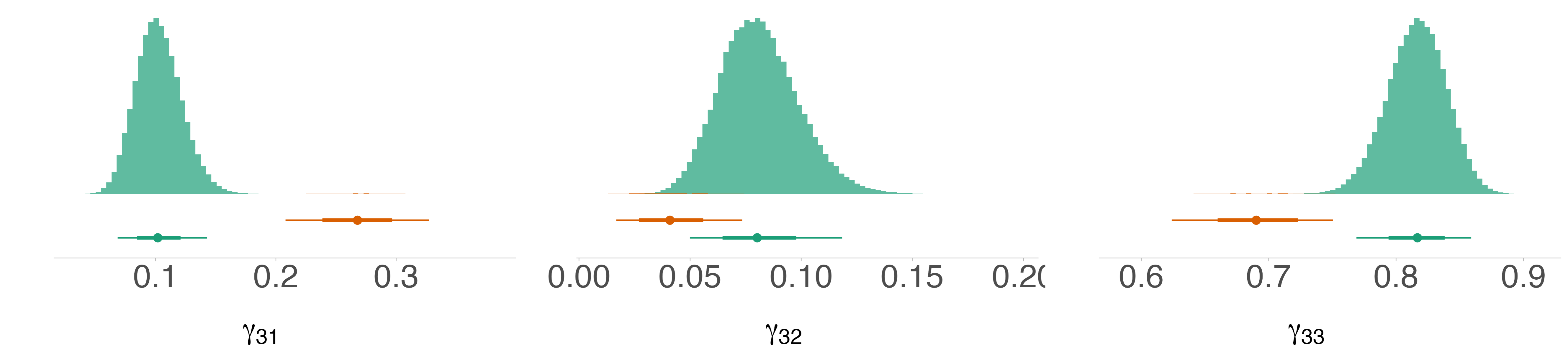}
\end{subfigure}
\caption{Histograms of the marginal posterior distributions estimated from the coldest chain. Row a) corresponds to the rate parameter $\lambda_n$ associated with the number of lunges data stream for each hidden state $n$, while rows b)—d) correspond to the transition probabilities $\gamma_{ij}$. The color indicates the mode to which the high-density region corresponds. The lines below the histograms are the mode-wise 95\% credible interval. The dot in the line indicates the posterior median, whereas the thicker line inside the 95\% credible interval indicates the mode-wise 66\% credible interval.}
\label{fig:sddp_lunges_cov}
\end{figure}

\begin{figure}[htp!]
%\vspace*{-5mm}
%\vspace*{-7mm}
\begin{subfigure}{1.0\textwidth}
  \centering
  b)
  \includegraphics[trim={0 0 0 0cm},clip,width=1\linewidth]{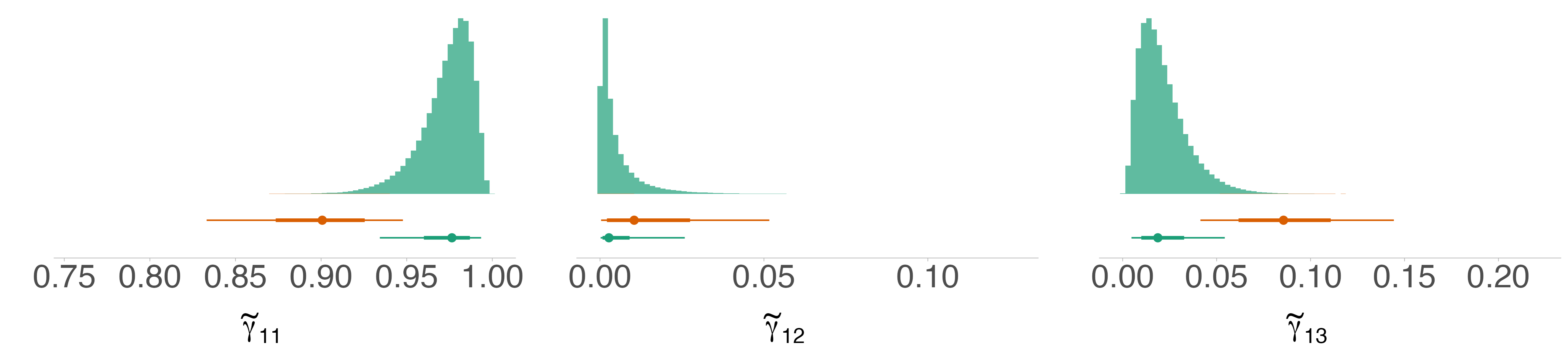}
\end{subfigure}

\vspace*{5mm}
\begin{subfigure}{1.0\textwidth}
  \centering
  c)
  \includegraphics[trim={0 0 0 0cm},clip,width=1\linewidth]{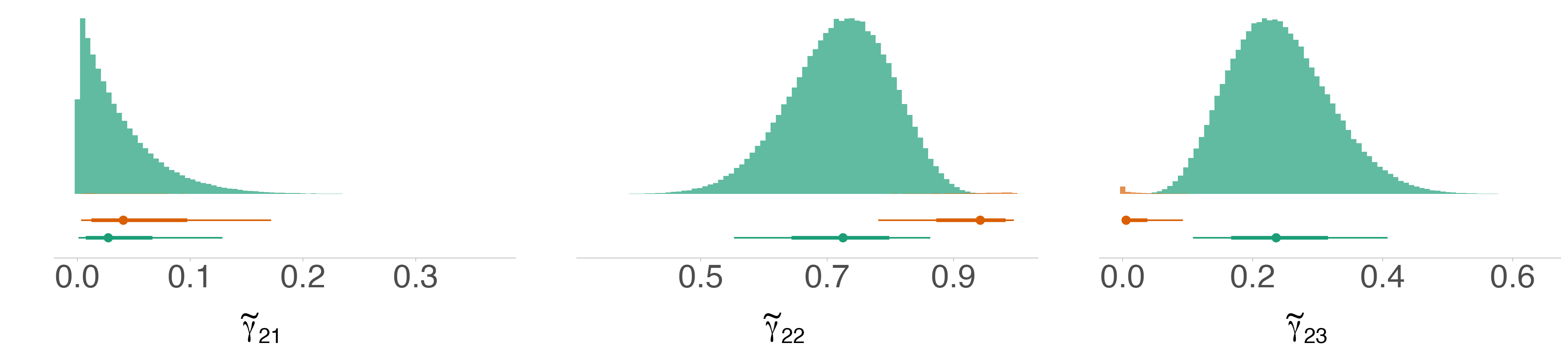}
\end{subfigure}

\vspace*{5mm}
\begin{subfigure}{1.0\textwidth}
  \centering
  d)
  \includegraphics[trim={0 0 0 0cm},clip,width=1\linewidth]{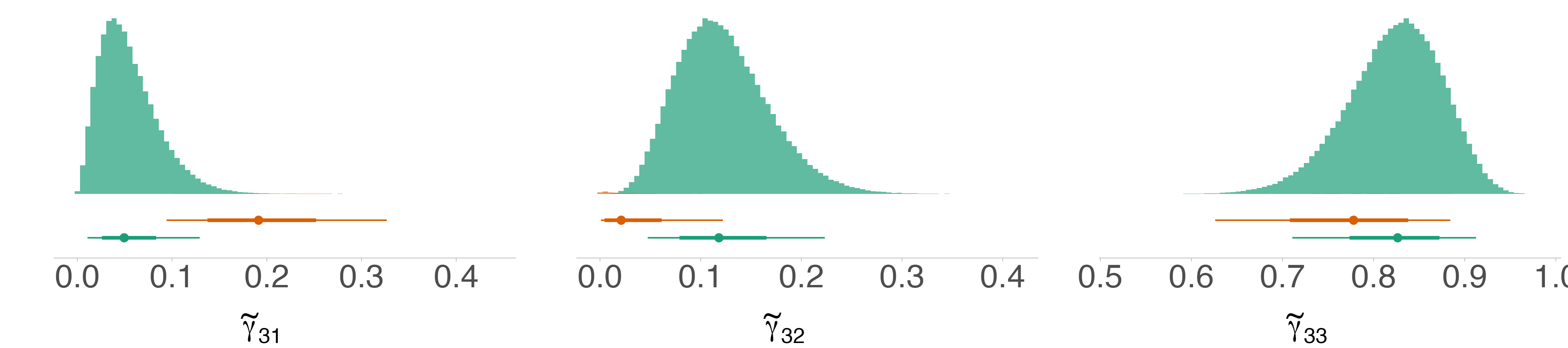}
\end{subfigure}
\caption{Histograms of the marginal posterior distributions of the transition probabilities in the presence of sound stimuli, estimated from the coldest chain related to the PT algorithm implementation with ID 1. The color indicates the mode to which the high-density region correspond. The lines below the histograms are the mode-wise 95\% credible interval. The dot in the line indicates the posterior median, whereas the thicker line inside the 95\% credible interval indicates the mode-wise 66\% credible interval.}
\label{fig:sddp_gamma_betas_cov}
\end{figure}

Following the same procedure as for the baseline 3-state HMM, the chain states from the coldest chain corresponding to the PT algorithm implementation with ID 1 were extracted for the computation of the marginal posterior distributions. Figure \ref{fig:sddp_lunges_cov} illustrates the same marginal posterior distributions shown in Figure \ref{fig:sddp_lunges}, but for the extended baseline 3-state HMM. Additionally, Figure \ref{fig:sddp_gamma_betas_cov} shows the marginal posterior distribution of the transition probabilities when sound stimuli is present.

\subsection{Results}

For the baseline 3-state HMM, the 95\% credible intervals of approximately two thirds of the estimated marginal state-dependent distribution parameters do not overlap. The parameters for which the 95\% credible intervals overlap are $\mu_{22}, \sigma_{23}, \mu_{31}, \sigma_{31}, \mu_{32}, \sigma_{32}, \sigma_{33}, \mu_{41}, \sigma_{41}, \sigma_{52}$. For the transition probabilities, the 95\% credible intervals for $\gamma_{22}$ and $\gamma_{23}$ do not overlap, while those for $\gamma_{11}$ overlap slightly. For the initial state probability estimates, all 95\% credible intervals overlap. See Table \ref{tab:results_baseline} for the posterior median parameter estimates and mode-wise 95\% credible intervals for these parameters. For the extended 3-state model with covariates, again approximately two thirds of the 95\% credible intervals for the state-dependent distribution parameters overlap across the two high-density regions, and these correspond to the same parameters as in the baseline model without covariates. For the entries of the transition probability matrix, when sound stimuli is absent, the 95\% credible intervals for $\gamma_{22}$ and $\gamma_{23}$ do not overlap, and the same is true for $\gamma_{31}$ and $\gamma_{33}$. When sound stimuli is present, the credible intervals for $\gamma_{33}$ across the two high-density regions do not overlap. See Appendix \ref{appdx:posterior_estimates_cov} for the posterior median parameter estimates and mode-wise 95\% credible intervals for these parameters.

\begin{table}[htp!]
\centering
\caption{Posterior median parameter estimates and 95\% credible intervals (in parentheses) computed mode-wise from samples of the PT algorithm implementation ID 1 for the baseline 3-state HMM. Results are organized by state-dependent parameters, transition probabilities, initial state distribution, and baseline working parameters corresponding to the reparameterization introduced in Equation \ref{eq:tpm_cov}.}
\label{tab:results_baseline}
\subcaption*{State-dependent parameters}
\scalebox{.7}{
\centering
\begin{tabular}{lcccccc}
\toprule
 & \multicolumn{2}{c}{State $n = 1$} & \multicolumn{2}{c}{State $n = 2$} & \multicolumn{2}{c}{State $n = 3$} \\
Variable & Mode $A$ & Mode $B$ & Mode $A$ & Mode $B$ & Mode $A$ & Mode $B$ \\
\midrule
$\lambda_{n}$ & 0.653 (0.579,0.733) & 0.428 (0.375,0.485) & 0.037 (0.004,0.098) & 2.817 (2.573,3.072) & 3.405 (3.217,3.601) & 2.92 (2.7,3.158) \\
$\mu_{2n}$ & 150 (142,159) & 182 (171,193) & 357 (324,395) & 404 (385,421) & 523 (510,536) & 569 (553,585) \\
$\sigma_{2n}$ & 89 (81,97) & 131 (120,143) & 224 (196,259) & 120 (107,134) & 124 (115,135) & 124 (113,137) \\
$\mu_{3n}$ & 68 (63,75) & 72 (66,78) & 96 (90,102) & 106 (101,111) & 154 (147,162) & 181 (172,191) \\
$\sigma_{3n}$ & 66 (60,74) & 67 (61,74) & 37 (32,43) & 38 (35,43) & 70 (64,77) & 69 (62,79) \\
$\mu_{4n}$ & 33 (30,35) & 30 (28,33) & 80 (69,93) & 112 (108,116) & 173 (167,179) & 216 (209,222) \\
$\sigma_{4n}$ & 24 (22,27) & 22 (20,25) & 73 (61,88) & 26 (23,29) & 60 (55,65) & 41 (37,46) \\
$\mu_{5n}$ & 207 (190,224) & 314 (291,339) & 712 (666,763) & 354 (315,396) & 414 (383,448) & 524 (475,580) \\
$\sigma_{5n}$ & 151 (134,169) & 268 (244,295) & 288 (255,329) & 235 (198,281) & 293 (264,326) & 380 (333,437) \\
\bottomrule
\end{tabular}

}
%\vspace{0.1cm}
\bigskip
\subcaption*{Transition probabilities and initial state distribution}
\scalebox{.7}{
\begin{tabular}{lcccccc}
\toprule
 & \multicolumn{2}{c}{State $n = 1$} & \multicolumn{2}{c}{State $n = 2$} & \multicolumn{2}{c}{State $n = 3$} \\
Variable & Mode $A$ & Mode $B$ & Mode $A$ & Mode $B$ & Mode $A$ & Mode $B$ \\
\midrule
$\gamma_{1n}$ & 0.913 (0.883,0.939) & 0.869 (0.838,0.896) & 0.023 (0.01,0.044) & 0.024 (0.011,0.042) & 0.062 (0.041,0.089) & 0.107 (0.082,0.135) \\
$\gamma_{2n}$ & 0.051 (0.019,0.099) & 0.064 (0.035,0.102) & 0.747 (0.671,0.816) & 0.925 (0.883,0.956) & 0.2 (0.137,0.273) & 0.01 (0.001,0.031) \\
$\gamma_{3n}$ & 0.09 (0.06,0.126) & 0.255 (0.2,0.317) & 0.088 (0.058,0.124) & 0.034 (0.013,0.067) & 0.821 (0.778,0.86) & 0.709 (0.646,0.767) \\
$\delta_{n}$ & 0.179 (0.08,0.323) & 0.271 (0.15,0.421) & 0.17 (0.065,0.318) & 0.224 (0.109,0.377) & 0.641 (0.477,0.788) & 0.495 (0.341,0.651) \\
\bottomrule
\end{tabular}

}
\bigskip
\subcaption*{Baseline working parameters}
\scalebox{1}{    \begin{tabular}{lcc}
\toprule
Variable & Mode $A$ & Mode $B$ \\
\midrule
$\alpha_{12}$ & -3.664 (-4.555,-3.02) & -3.593 (-4.368,-3.003) \\
$\alpha_{13}$ & -2.685 (-3.128,-2.304) & -2.094 (-2.387,-1.827) \\
$\alpha_{21}$ & -2.683 (-3.694,-1.97) & -2.667 (-3.293,-2.158) \\
$\alpha_{23}$ & -1.321 (-1.771,-0.912) & -4.578 (-6.755,-3.366) \\
$\alpha_{31}$ & -2.213 (-2.64,-1.84) & -1.021 (-1.339,-0.718) \\
$\alpha_{32}$ & -2.234 (-2.674,-1.854) & -3.029 (-3.984,-2.327) \\
\bottomrule
\end{tabular}

}
\end{table}
 
\subsection{Comparing uncertainty quantification results in the presence of multiple modes}

As mentioned in Section \ref{subsec:blue_whale_data}, the blue whale dive data were previously analyzed in \cite{deruiter_multivariate_2017} with inference conducted using maximum likelihood estimation. To quantify uncertainty around the point estimates, they constructed Wald-based intervals. In that study, seven data streams were used for the observable process, whereas in our analysis we used five of the seven data streams. Their baseline model was a 3-state HMM with the same assumptions as the baseline model used in this paper. During model fitting, they reported the presence of multiple local maxima.

The estimates of the state-dependent distribution parameters reported in \citep{deruiter_multivariate_2017} were generally captured within one of the two high-density regions identified for the baseline 3-state HMM fitted here, specifically for the mode with the highest estimated weight, mode $A$, with the exception of five parameters: $\mu_{21}$, $\mu_{32}$, $\sigma_{32}$, $\mu_{42}$, and $\sigma_{51}$. For the entries of the transition probability matrix, the credible intervals of the same mode as the state-dependent distribution parameters, mode $A$, captured all of the estimates presented in \citep{deruiter_multivariate_2017}. For reference, the results from \citep{deruiter_multivariate_2017} were reproduced following their procedure and can be found in Appendix \ref{appdx:hmm_mle}. Because the number of data streams differ between our baseline model and that of \citep{deruiter_multivariate_2017}, differences in parameter estimates were expected to occur. However, the uncertainty associated with the estimates from one of the two modes we found is not reported in \cite{deruiter_multivariate_2017}, leading to different conclusions between their work and ours, namely, the mode with the lowest estimated weight, mode $B$. 

Comparing the results from \cite{deruiter_multivariate_2017} and our full Bayesian analysis highlights the difficulties with quantifying uncertainty for our parameters in the presence of a multimodal likelihood, and subsequently posterior distribution. In a frequentist framework, \cite{chen_statistical_2023} highlights how different confidence intervals can be constructed depending on the choice of interval estimation. However, a full Bayesian analysis is quite straightforward as we are only interested in computing or sampling from the joint posterior distribution, irrespective of how many modes it may have. To construct credible intervals from the joint posterior distribution, we can use the estimated weights for each mode, together with the uncertainty within each mode. 

For the blue whale movement data, we consider the average of the estimated weights across all PT algorithm implementations, rather than relying on a single run. Specifically, we take the estimated weight of mode $A$ to be 0.819 and of mode $B$ as 0.181. From here, we ensure to construct credible intervals from the marginal distributions as a mixture of mode-specific credible intervals that are weighted by either 0.819 if belonging to mode A or 0.181 if belonging to mode B. Interpretation of these intervals in a Bayesian setting is also straightforward as making probabilistic statements over disjoint intervals need not be different than if only a single mode existed. The intuitive construction of credible intervals per mode to account for uncertainty, together with the estimation of mode weights to assign probabilities to each interval, highlights an advantage of the Bayesian framework over frequentist approaches. In particular, it provides a natural way to construct credible intervals within each high-density region, thereby accounting for the full uncertainty in the parameter space, which can be directly interpreted since uncertainty is expressed as a distribution.

These differences have important implications for inference. If posterior summaries are based on a single mode, conclusions may differ substantially. Using posterior medians from mode $A$, an observation generated from hidden state 1 would be characterized by a mean number of lunges of 0.653, a mean dive duration of 150 seconds, a mean post-dive surface duration of 68 seconds, a mean maximum depth of 33 meters, and a mean step length of 207 meters. In contrast, using mode $B$, the same hidden state would correspond to a mean number of lunges of 0.428, which is an average decrease of 0.225 units compared to mode $A$, a mean dive duration of 182 seconds, an average increase of 32 seconds, a mean post-dive surface duration of 72 seconds, an average increase of 4 seconds, a mean maximum depth of 30 meters, an average decrease of 3 meters, and a step length of 314 meters, which is an average increase of 107 meters compared to mode $A$. Similar discrepancies arise for results in hidden state 2 and 3. In contrast, our Bayesian credible intervals provide that state 1 is characterized by a bimodal distribution in which 0.819 percent of our lunges are generated from a distribution with estimated mean 0.653 and 0.181 percent from a distribution with estimated 0.428, with uncertainty intervals provided easily from the joint posterior distribution. 

\section{Discussion}\label{sec:discussion}

In this work, we provide implementation guidelines for applying the parallel tempering (PT) algorithm to hidden Markov models (HMMs) within a Bayesian framework, with the goal of conducting inference in the presence of genuine multimodality in the posterior distribution. For the PT algorithm implementation, a temperature schedule targeting a 0.23 swap acceptance rate was proposed; the tempered replicas were defined as power-posterior versions of the target distributions, and within-temperature moves for marginal exploration are proposed to be performed using a component-wise Metropolis–Hastings algorithm. A new prior formulation was also introduced for the working parameters $\alpha_{0}^{(ij)}$ and the coefficients $\alpha_{1l}^{(ij)}$ associated when incorporating a categorical covariate into the transition probability matrix using Equation \ref{eq:tpm_cov} in Section \ref{sec:hmm}, inducing a uniform distribution on the simplex corresponding to the rows of the transition probability matrix. Following these implementation guidelines, we applied the PT algorithm to ecological time-series data consisting of blue whale dives from a group of blue whales that were exposed to sound stimuli. For each dive, five variables summarizing blue whale movement were recorded: number of lunges, dive duration, post-dive surface duration, maximum depth, and step length. These data streams were modeled using a baseline 3-state HMM and an extended version that incorporated the occurrence of sound stimuli in the entries of the transition probability matrix. For each model, we conducted 10 independent PT algorithm implementations. Consistent swap acceptance rates across runs, along with information from the coldest replica moving back and forth to the hottest replica, provided evidence of a successful implementation. Label switching correction was conducted in a post-sampling process for the coldest chains for all PT algorithm implementations, revealing genuine multimodality in the joint posterior distribution. Specifically, two high-density regions were identified for both models. Consistent running weight estimates across PT algorithm implementations further indicated that the PT algorithm successfully explored both modes.

The prior distribution introduced in Section \ref{subsec:bayes_hmm} can be extended to multiple exogenous categorical covariates; however the number of parameters to estimate does not increase linearly. Instead, it grows with the total number of combinations of outcomes across the multiple categorical covariates. Specifically, if we have $C$ categorical covariates $\{Z^{(1)}_{t} \},\ldots, \{Z^{(C)}_{t} \}$ with corresponding number of outcomes $O_1,\ldots, O_C$, then maintaining the provided Gumbel prior requires constructing a new latent categorical covariate whose values correspond to all possible combinations of the outcomes. This results in $\prod_{i=1}^{C} O_i$ outcomes for the new covariate, rather than $\sum_{i=1}^{C} O_i$, which would be the case when using popular prior assumption approaches for covariates.

The inverse temperature schedules were constructed to target a swap acceptance rate of 0.234 across all adjacent temperatures. While the optimal inverse temperature schedule conveys constant swap acceptance rates, it has not been verified theoretically that 0.234 is the correspond to the value for the optimal solution in the case of hidden Markov models. Consequently, the optimal swap acceptance rate for hidden Markov models remains an open research question.

Artificial identifiability constraints, that is, ordering constraints, were used to correct for label switching in a post-sampling process. Since the high-density regions were well separated, this approach was suitable for our models. However, this situation is rare. \citep{jasra_markov_2005} highlights that finding appropriate identifiability constraints in multivariate problems is often nearly impossible. There are cases in which the posterior distribution is genuinely multimodal, and no identifiability constraints can successfully isolate both major and minor modes (see, e.g., \cite{grun_dealing_2009}). There is an extensive literature on relabeling methods for addressing label switching. Classical approaches include label-invariant loss functions \cite{celeux_computational_2000}, the pivotal reordering algorithm \cite{marin_bayesian_2005}, Kullback–Leibler divergence–based algorithms \cite{stephens_dealing_2000}, and the equivalence classes representatives (ECR) algorithm \cite{papastamoulis_artificial_2010}. More recent approaches are based on optimal transport \cite{monteiller_alleviating_2019}. An extensive review of artificial identifiability constraints, relabeling algorithms, and label-invariant loss functions is provided in \cite{jasra_markov_2005}. Additionally, \cite{papastamoulis_labelswitching_2016} compiles several of these methods into the R package label.switching, which can be used to assess whether there are differences compared to the artificial ordering constraints applied in this problem.

For the model with covariates, the weight of one of the two modes is very small, ranging between 0.00268 and .00482 across all PT algorithm implementations. Nevertheless, we believe this highlights a strength of the parallel tempering algorithm, which is its ability to capture uncertainty across all high-density regions, even those with relatively small weights. The results for both the baseline and extended model demonstrate that accounting for uncertainty associated with either local or global modes can lead to substantially different uncertainty quantification, with implications for inference.

Our results show that accounting for the uncertainty associated with multiple high-density regions in the posterior parameter space, when present, can lead to substantially different parameter estimates when summarizing the posterior within each mode independently. Furthermore, since uncertainty is represented as a distribution in a Bayesian framework, this allows us to account for uncertainty across all high-density regions by assigning each mode a weight using the running weight estimates introduced in Section \ref{subsec:genuine_multimodality}. This highlights an advantage of the Bayesian approach over frequentist methods, as we can construct credible intervals within each high-density region, thereby accounting for the full uncertainty in the parameter space. This reinforces our belief that genuine multimodality in the joint posterior distribution should be treated as an inferential challenge in addition to a computational one. Nonetheless, it is important to note that the uncertainty captured by local maxima does not necessarily correspond to meaningful information for inference. Model validation is essential to determine whether such multimodality arises as a computational artifact or reflects substantively relevant uncertainty. The development of diagnostic and validation methods for HMMs under genuine multimodality is a crucial next step and a potential direction for future research.

\subsection*{Acknowledgements}

During this project, Marco A. Gallegos-Herrada received funding from the Secretaría de Ciencia, Humanidades, Tecnología e Innovación (SECIHTI). He also received support from the Canadian Statistical Sciences Institute (CANSSI). Vianey Leos-Barajas and Jeffrey S. Rosenthal were financially supported by the Natural Sciences and Engineering Research Council of Canada. Computational resources were provided by the Digital Research Alliance of Canada (DRAC). We give thanks to Stacy De Ruiter for many conversations that proved very useful to development of the paper. 

\newpage
\bibliographystyle{plainnat}
\bibliography{references_zotero}

\clearpage

\appendix

\section{Appendix}

\subsection{Derivation of priors for $\alpha_{0}^{(ij)}$ and $\alpha_{1l}^{(ij)} $}\label{appdx:priors_derivation}

The rationale for constructing the priors for the components $\alpha_{0}^{(ij)}$ presented in Section \ref{subsec:bayes_hmm} is as follows. If we consider
\[
\gamma_{ij} = \dfrac{-\log(U_{ij})}{\sum_{k=1}^{N}-\log(U_{ik})}, \quad j \in \{1, \ldots , N\},
\]
with $U_{ij} \sim U(0,1)$, we have that $(\gamma_{i1}, \ldots, \gamma_{iN})$ follows a $\text{Dirichlet}(\boldsymbol{1}_N)$ distribution. Let $\zeta_{ij} = \log(-\log(U_{ij}))$. Then, we have
\begin{equation}
    \log\left(\dfrac{\gamma_{ij}}{\gamma_{ii}}\right) = \zeta_{ij} - \zeta_{ii}, \quad i\neq j.    
\end{equation}
From the reparameterization of the transition probabilities shown in Equation \ref{eq:tpm_cov}, we have that $\alpha_{0}^{(ij)} = \zeta_{ij} - \zeta_{ii}$ and $-\zeta_{ij} \sim \text{Gumbel}(0,1) ,\, i \neq j$. Assuming $\zeta_{ii} \perp \zeta_{ij}$, $i\neq j$, we have
\begin{align*}
-\alpha_{ij}\mid \zeta_{ii} &\sim \text{Gumbel}( \zeta_{ii}, 1),\\
 -\zeta_{ii} &\sim \text{Gumbel}(0,1),
\end{align*}
which leads to $(\gamma_{i1}, \ldots, \gamma_{iN}) \sim \text{Dirichlet}(\boldsymbol{1}_N)$, for $i \in \{1, \ldots, N\}$. A consequence this prior setting is that the marginal distribution of $\alpha_{0}^{(ij)}$ is Logistic$(0,1)$, with marginal distributions of $\alpha_{0}^{(ij)}$ correlated. The dependence on the latent component $\zeta_{ii}$ guarantees identifiability, as this induces a one-to-one transformation between the baseline components and the transition probabilities in the original scale. 

To construct a prior that induces equal weights over the simplex corresponding to the transition probability rows when a categorical covariate $z_t$ is incorporated, we adapt the prior formulation from Equation \ref{eq:tpm_cov} to the categorical case described in Equation \ref{eq:tpm_cov_categorical}. Specifically, we leverage the prior specification for $\alpha_{0}^{(ij)}$. Given the construction of priors for the parameters $\alpha_{0}^{(ij)}$, we define
\[
\alpha_{1l}^{(ij)} = \zeta_{ij}^{(l)} - \zeta_{ij},
\]
with $-\zeta_{ij}^{(l)} \sim \text{Gumbel}(0,1)$. This implies 
\begin{align*}
\alpha_{0}^{(ij)} + \alpha_{1l}^{(ij)} = \zeta_{ij}^{(l)} - \zeta_{ii} \sim \text{Logistic}(0,1),\\
    -(\alpha_{0}^{(ij)} + \alpha_{1l}^{(ij)})\mid \zeta_{ii} \sim \text{Gumbel}(\zeta_{ii},1),\\
    -\alpha_{1l}^{(ij)}\mid \alpha_{0}^{(ij)}, \zeta_{ii} \sim \text{Gumbel}(\alpha_{0}^{(ij)} +\zeta_{ii}, 1).
\end{align*}

\clearpage

\subsection{PT diagnostics}\label{appdx:diagnostics}

Key metrics for assessing the effectiveness of the PT algorithm include the swap acceptance rate between adjacent temperatures and the occurrence of round trips, where information from the coldest replica reaches the hottest replica and returns to the coldest. The latter indicates how effectively information is propagated across replicas. To evaluate round trips, trace plots can be used to visually assess the movement of information across the tempered replicas and identify any issues. Below, we present trace plots of the coldest replica moving across all tempered replicas for both the baseline 3-state HMM and its extension with covariates in the transition probabilities for all the PT algorithm implementations.

\subsubsection*{Baseline 3-state HMM}
    
\begin{figure}[h!]
    \centering
        \includegraphics[trim={0 0 3.5cm 0cm},clip, scale=.076]{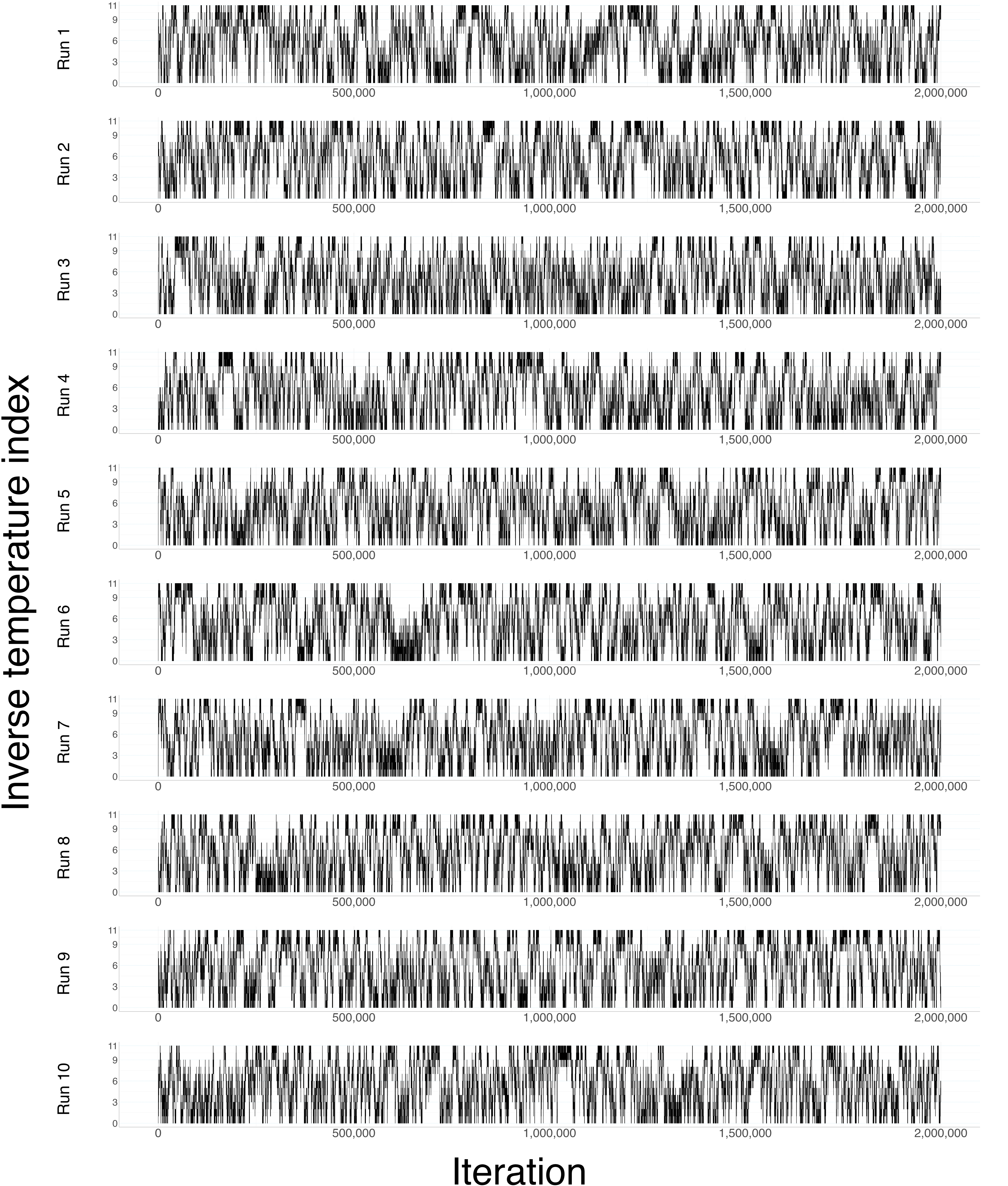}
    \caption{Traceplot of information from the coldest replica moving across all tempered replicas for the 10 PT algorithm implementations of the baseline 3-state HMM.}
    \label{fig:traceplot_baseline}
\end{figure} 
%\vspace{-5cm}

\subsubsection*{3-state HMM with sound stimuli covariate}

\begin{figure}[h!]
    \centering
        \includegraphics[trim={0 0 3.5cm 0cm},clip, scale=.076]{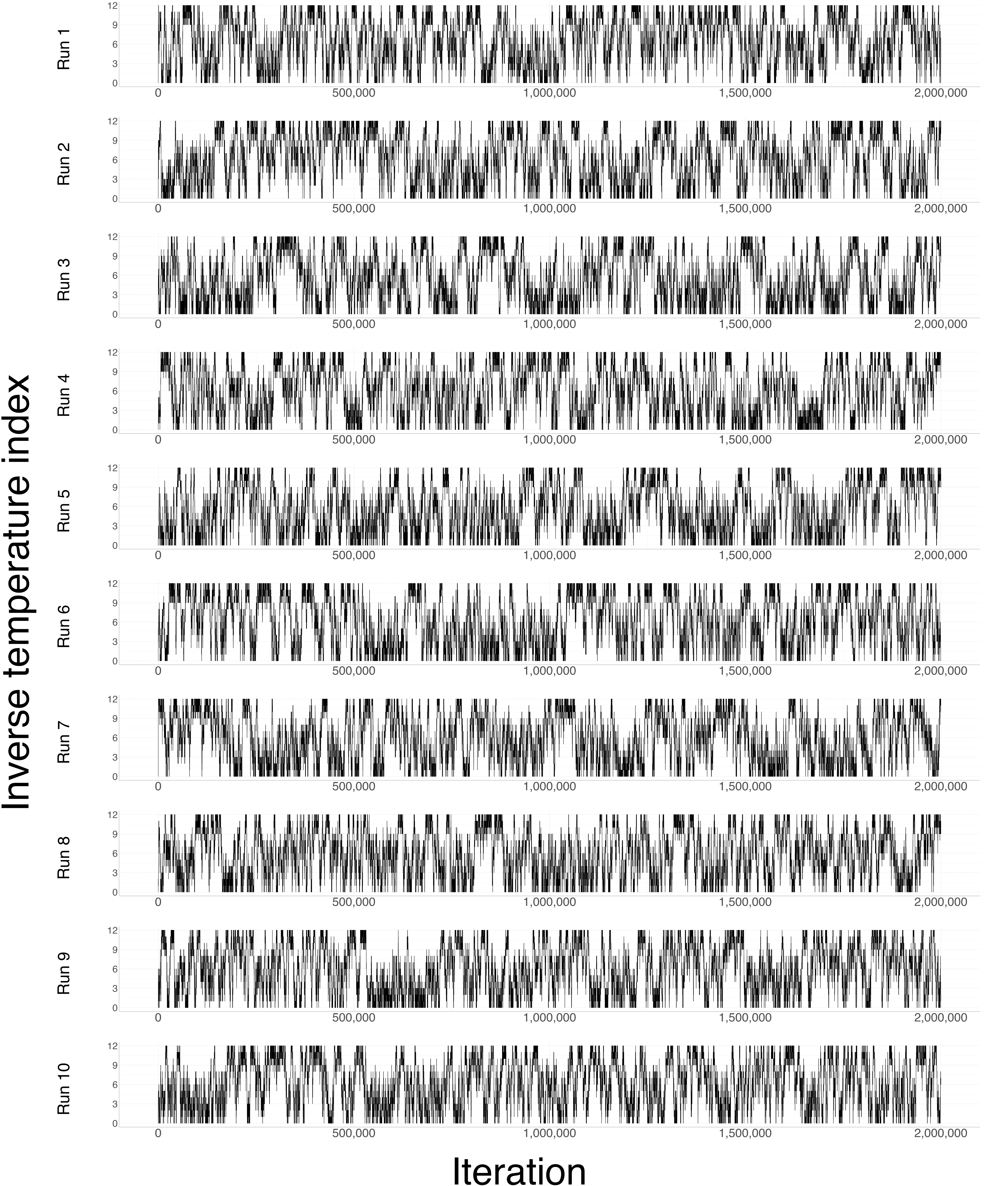}
    \caption{Traceplot of information from the coldest replica moving across all tempered replicas for the 10 PT algorithm implementations of the extended 3-state HMM with covariates.}
    \label{fig:traceplot_extended_cov}
\end{figure}  
%\vspace{-10cm}

\clearpage
\subsection{Pseudocode for implementing CWMH for HMMs}\label{appdx:cwmh}

The following pseudocode outlines the implementation of the Component-Wise Metropolis–Hastings (CWMH) algorithm for sampling from an $N$-state HMM with a single $P$-dimensional observation sequence, with the option to incorporate a binary covariate. Although the algorithm is presented for a single observation sequence, it can be readily generalized by following the Algorithm \ref{algthm:power_posterior} in Appendix \ref{appdx:logLik_powerPosterior_algorithms}. Similarly, the algorithm can be extended to accommodate a covariate with $L$ categories by iterating over the $L-1$ levels, since one level can be absorbed into the baseline working parameter.

\begin{algorithm}[H]
\caption{CWMH($P$-dimensional observation sequence $\boldsymbol{y_{1:T}}$, 
working parameters $( \alpha_{0:1}^{(ij)})_{ij}$, initial state distribution $\boldsymbol{\delta}$, state-dependent parameters $\boldsymbol{\phi} = (\boldsymbol{\phi_1} ,\ldots, \boldsymbol{\phi_P} )$, covariate indicator $c$)}
\begin{algorithmic}[1]

\Require A proposal mechanism $q(\cdot, \cdot)$ for each sub-block of parameters.

\vspace{0.2cm}

\State includeCovariate $\gets$ FALSE
\If{$c == 1$}
    \State includeCovariate $\gets$ TRUE
\EndIf

\vspace{0.2cm}
\Statex \textbf{1. Update baseline working parameters}

\For{$i = 1, \dots, N$}
    \State Propose $\left(\alpha_{0}^{(ij)}\right)_{j}^{*}
    \sim q\left(\left(\alpha_{0}^{(ij)}\right)_j , \cdot \right)$
    
        \State $\boldsymbol{\theta} \gets \left(\left(\alpha_{0}^{(ij)} \right)^{*}_1 ,\ldots,\left(\alpha_{0}^{(ij)}\right)_{i}, \ldots \left(\alpha_{0}^{(ij)} \right)_{N}, \boldsymbol{\delta}, \boldsymbol{\alpha_{1}^{(ij)}}, \boldsymbol{\phi} \right)$

    \State $\boldsymbol{\theta}^{*} \gets \left(\left(\alpha_{0}^{(ij)} \right)^{*}_1 ,\ldots,\left(\alpha_{0}^{(ij)}\right)_{i}^{*}, \ldots \left(\alpha_{0}^{(ij)} \right)_{N}, \boldsymbol{\delta}, \boldsymbol{\alpha_{1}^{(ij)}}, \boldsymbol{\phi} \right)$

    \State Compute:
    \[
    A = \log p(\theta^* \mid \boldsymbol{y_{1:T}}) - \log p(\theta \mid \boldsymbol{y_{1:T}})
    + \log q\left(\left(\alpha_{0}^{(ij)}\right)^{*}_j , \left(\alpha_{0}^{(ij)}\right)_j \right) - \log q\left(\left(\alpha_{0}^{(ij)}\right)_j , \left(\alpha_{0}^{(ij)}\right)^{*}_j \right) 
    \]
        %\vspace{0.2cm}
      \State $U \sim \text{Unif}(0,1)$
      \If{$\log(U) > A$}
        \State $\left(\alpha_{0}^{(ij)}\right)^{*}_{j} \gets \left(\alpha_{0}^{(ij)}\right)_{j}$
    
      \EndIf

    \EndFor

    \State Let $\left(\boldsymbol{\alpha_{0}^{(ij)}}\right)^{*} \gets \left(\left(\alpha_{0}^{(ij)}\right)^{*}\right)_{ij}$
    \vspace{0.2cm}

    \vspace{0.2cm}

    \State $U \sim \text{Unif}(0,1)$
      \If{$\log(U) > A$}
        %\State swap $x^{(k)} \leftrightarrow x^{(k+1)}$
        \State $\boldsymbol{\delta}^{*} \gets \boldsymbol{\delta}$
    
      \EndIf

    \vspace{0.2cm}
    
    \Statex \textbf{2. Update initial distribution}

    \State Propose $\boldsymbol{\delta}^* \sim q(\boldsymbol{\delta}, \cdot)$

    \State $\boldsymbol{\theta}^{*} \gets \left(\left(\boldsymbol{\alpha_{0}^{(ij)}} \right)^{*}, \boldsymbol{\delta}, \boldsymbol{\alpha_{1}^{(ij)}}, \boldsymbol{\phi} \right)$
    \State $\boldsymbol{\theta}^{*} \gets \left(\left(\boldsymbol{\alpha_{0}^{(ij)}} \right)^{*}, \boldsymbol{\delta}^{*}, \boldsymbol{\alpha_{1}^{(ij)}}, \boldsymbol{\phi} \right)$        

\algstore{cwmh_alg}

\end{algorithmic}
\end{algorithm}

\begin{algorithm}
    \begin{algorithmic}[1]

    \algrestore{cwmh_alg}

    \State Compute:
    \[
    A = \log p(\theta^* \mid \boldsymbol{y_{1:T}} ) - \log p(\theta \mid \boldsymbol{y_{1:T}} )
    + \log \frac{q(\boldsymbol{\delta}^{*}, \boldsymbol{\delta})}{q(\boldsymbol{\delta} , \boldsymbol{\delta}^{*})}
    \]

        \Statex \textbf{3. Update covariate parameters (if applicable)}

    \If{includeCovariate}
        \For{$i = 1, \dots, N$}
            \State Propose $\left(\alpha_{1}^{(ij)}\right)_{j}^{*}
            \sim q\left(\left(\alpha_{1}^{(ij)}\right)_j , \cdot \right)$
        
            \State $\boldsymbol{\theta} \gets \left(\left(\boldsymbol{\alpha_{0}^{(ij)}} \right)^{*}, \boldsymbol{\delta}^{*} , \left(\alpha_{1}^{(ij)} \right)^{*}_1 ,\ldots,\left(\alpha_{1}^{(ij)}\right)_{i}, \ldots \left(\alpha_{1}^{(ij)} \right)_{N} , \boldsymbol{\phi} \right)$
            \State $\boldsymbol{\theta}^{*} \gets \left(\left(\boldsymbol{\alpha_{0}^{(ij)}} \right)^{*}, \boldsymbol{\delta}^{*} , \left(\alpha_{1}^{(ij)} \right)^{*}_1 ,\ldots,\left(\alpha_{1}^{(ij)}\right)_{i}^{*}, \ldots \left(\alpha_{1}^{(ij)} \right)_{N} , \boldsymbol{\phi} \right)$

                    \State Compute:
            \[
            A = \log p(\boldsymbol{\theta}^{*} \mid \boldsymbol{y_{1:T}}) - \log p(\boldsymbol{\theta} \mid \boldsymbol{y_{1:T}})
            + \log \frac{q\left(\left(\alpha_{1}^{(ij)}\right)^{*}_j , \left(\alpha_{0}^{(ij)}\right)_j \right)}{q\left(\left(\alpha_{1}^{(ij)}\right)_j , \left(\alpha_{1}^{(ij)}\right)^{*}_j \right)}
            \]
                \vspace{0.2cm}
              \State $U \sim \text{Unif}(0,1)$
              \If{$\log(U) > A$}
                %\State swap $x^{(k)} \leftrightarrow x^{(k+1)}$
                \State $\left(\alpha_{1}^{(ij)}\right)^{*}_{j} \gets \left(\alpha_{1}^{(ij)}\right)_{j}$
            
              \EndIf
                %\State Accept $\Gamma_i^*$ with probability $\min(1, e^{\log \alpha})$
            \EndFor
        
            \State Let $\left(\boldsymbol{\alpha_{1}^{(ij)}}\right)^{*} \gets \left(\left(\alpha_{1}^{(ij)}\right)^{*}\right)_{ij}$
        
    \EndIf
    \vspace{0.2cm}
    \Statex \textbf{4. Update state-dependent parameters}

    \For{$p = 1, \dots, P$}
        \State Propose $\phi_p^* \sim q(\phi_{p}, \cdot )$

            \State $\boldsymbol{\theta} \gets \left(\left(\boldsymbol{\alpha_{0}^{(ij)}} \right)^{*}, \boldsymbol{\delta}^{*} , \left(\boldsymbol{\alpha_{1}^{(ij)} }\right)^{*}, \boldsymbol{\phi}^{*}_{1}, \ldots , \boldsymbol{\phi}_{p}, \ldots , \boldsymbol{\phi}_{P} \right)$
            \State $\boldsymbol{\theta}^{*} \gets \left(\left(\boldsymbol{\alpha_{0}^{(ij)}} \right)^{*}, \boldsymbol{\delta}^{*} , \left(\boldsymbol{\alpha_{1}^{(ij)} }\right)^{*}, \boldsymbol{\phi}^{*}_{1}, \ldots , \boldsymbol{\phi}^{*}_{p}, \ldots , \boldsymbol{\phi}_{P} \right)$
        \State Compute:
        \[
        A = \log p(\boldsymbol{\theta}^* \mid \boldsymbol{y_{1:T}}) - \log p(\boldsymbol{\theta} \mid \boldsymbol{y_{1:T}})
        + \log \frac{q(\phi^{*}_{k}, \phi_{k})}{q(\phi_{k}, \phi^{*}_{k})}
        \]
        \vspace{0.2cm}
      \State $U \sim \text{Unif}(0,1)$
      \If{$\log(U) > A$}
        \State $\phi_{p}^{*} \gets \phi_{p}$
      \EndIf

    \EndFor
    \State Let $\boldsymbol{\phi}^{*} \gets (\boldsymbol{\phi^{*}_{1}},\ldots,\boldsymbol{\phi^{*}_{P}}) $
    \vspace{0.2cm}

\vspace{0.2cm}
\State \Return $\left(\boldsymbol{\alpha_{0}^{(ij)}} \right)^{*}, \boldsymbol{\delta}^{*} , \left(\boldsymbol{\alpha_{1}^{(ij)} }\right)^{*}, \boldsymbol{\phi}^{*} $
    \end{algorithmic}
\end{algorithm}

\subsection{Log-likelihood computation for HMMs via the forward algorithm and log-power posteriors}\label{appdx:logLik_powerPosterior_algorithms}

The algorithm \ref{algthm:log_likelihood} provides pseudocode for computing the log-likelihood of an $N$-state HMM using the forward algorithm \citep{zucchini_hidden_2017}, with the option to incorporate a binary covariate. The algorithm is presented for a single observation sequence; however, it can be applied independently to multiple sequences, in which case the log-likelihood is obtained by summing the contributions from each sequence. Additionally, it can be extended to accommodate a categorical covariate with $L$ possible outcomes.

Algorithm \ref{algthm:power_posterior} shows how to compute the log-power posterior under the same structure as Algorithm \ref{algthm:log_likelihood}, with the main difference being that it explicitly accounts for multiple independent observation sequences. In this case, the evaluation of the log-likelihood for multiple sequences refers to computing the log-likelihood for each sequence independently and summing the results.

\begin{algorithm}[H]
\caption{LogLikelihood(observations $\boldsymbol{y_{1:T}}$, latent process $z_{1:T} $, initial state distribution $\boldsymbol{\delta}$, working parameters $( \alpha_{0:1}^{(ij)})_{ij}$, state-dependent parameters $\boldsymbol{\phi}$, covariate indicator $c$)}
\begin{algorithmic}[1]

\State includeCovariate $\gets$ FALSE
\If{$c == 1$}
    \State includeCovariate $\gets$ TRUE
\EndIf

\vspace{0.2cm}

\For{$i = 1, \dots, N$}
    \Comment{Compute the forward variable at time $t=1$ in the log scale}
    \State $\psi_i(1) \gets \log(\delta_i) + \log[p(\boldsymbol{y_1} \mid S_1 = i, \boldsymbol{\phi})]$
\EndFor

\vspace{0.2cm}

\If{!includeCovariate}
    \For{$i = 1, \dots, N$}
        \Comment{Compute $i$-th row of t.p.m. without covariates using equation \ref{eq:tpm_cov}}
        \State $\Gamma_{i\,\cdot} \gets \text{multinomial-logit}((\alpha_{0:1}^{(ij)})_j, 0 )$
    \EndFor
    \EndIf

\vspace{0.2cm}

\For{$t = 2, \dots, T$}
            \Comment{Compute the forward probabilities at time $t>2$ in the log scale}
\If{includeCovariate}
    \For{$i=1,\ldots,N$}
        \Comment{Compute $i$-th row of t.p.m. with covariates using equation \ref{eq:tpm_cov}}
        \State $\Gamma_{i\,\cdot} \gets \text{multinomial-logit}((\alpha_{0:1}^{(ij)})_j, z_t )$
        \EndFor
\EndIf
    \For{$j = 1, \dots, N$}
        \State $b_j(t) \gets p(\boldsymbol{y_t} \mid S_t = j, \boldsymbol{\phi})$

        \State $\psi_j(t) \gets \text{log-sum-exp}\left[\psi_1(t-1) + \log(\Gamma_{1j}),\ldots , \psi_N(t-1) + \log(\Gamma_{Nj}) \right] + \log(b_j(t)) 
        $
    \EndFor
    
\EndFor

\State $\ell \gets \text{log-sum-exp}\left[\psi_1(T),\ldots , \psi_N(T) \right] $
\vspace{0.2cm} \Comment{Compute log-likelihood}

\vspace{0.2cm}
\State \Return $\ell$

\end{algorithmic}
\label{algthm:log_likelihood}
\end{algorithm}

\begin{algorithm}[H]
\caption{LogPowerPosterior(observations $\left(\boldsymbol{y^{(w)}_{1:T_w}}\right)_w$, latent process $\left(z^{(w)}_{1:T_w}\right)_w$, initial state distribution $\boldsymbol{\delta}$, working parameters $( \alpha_{0:1}^{(ij)})_{ij}$, state-dependent parameters $\boldsymbol{\phi}$, covariate indicator $c$, inverse temperature $\beta_m$)}
\begin{algorithmic}[1]

\Require Prior densities $p(\boldsymbol{\delta})$, $p\left(( \alpha_{0:1}^{(ij)})\right)$, $p(\boldsymbol{\phi})$

\vspace{0.2cm}

\State $\ell \gets 0$
\Comment{Initialize log-likelihood}
\For{$w=1,\ldots,W$}
\State $\ell \gets \ell +\text{LogLikelihood}
\left(\boldsymbol{y^{(w)}_{1:T_w}}, z^{(w)}_{1:T_w}, \boldsymbol{\delta}, (\alpha_{0:1}^{(ij)})_{ij} , \boldsymbol{\phi}, c\right)$

\Comment{Log-likelihood for sequence $w$}
\EndFor
\vspace{0.2cm}

\State $\ell_{\text{temp}} \gets \ell / \beta_m$
\Comment{Apply likelihood tempering}
\vspace{0.2cm}

\State $\ell_{\text{temp}} \gets \ell_{\text{temp}} + \log \left[ p(\boldsymbol{\delta})\right] + \log \left[p\left(( \alpha_{0:1}^{(ij)})\right)\right] + \log\left[p(\boldsymbol{\phi})\right]$
\Comment{Add log-priors (not tempered)}

\vspace{0.2cm}
\State \Return $\ell_{\text{temp}}$

\end{algorithmic}
\label{algthm:power_posterior}
\end{algorithm}

\subsection{Prior distributions}\label{appdx:priors_implementation}

The prior distributions for the state-dependent distribution parameters are as follows:\\

Number of lunges:
\[
\lambda_{n} \sim  \text{Gamma}(1.5,0.5) \quad n = 1, 2, 3
\]

Dive duration:
\vspace{-5mm}
\begin{align*}
    \mu_{2n} \sim  \text{Gamma}(3,.01) \quad n = 1, 2, 3\\
    \sigma_{2n} \sim \text{Gamma}(3,.01) \quad n = 1, 2, 3
\end{align*}

Surface duration:
\vspace{-5mm}
\begin{align*}
    \mu_{3n} \sim  \text{Gamma}(3,.01) \quad n = 1, 2, 3\\
    \sigma_{3n} \sim \text{Gamma}(3,.01) \quad n = 1, 2, 3
\end{align*}

Maximum depth:
\vspace{-5mm}
\begin{align*}
    \mu_{4n} \sim  \text{Gamma}(3,.01) \quad n = 1, 2, 3\\
    \sigma_{4n} \sim \text{Gamma}(3,.01) \quad n = 1, 2, 3
\end{align*}

Step length:
\vspace{-5mm}
\begin{align*}
    \mu_{5n} \sim  \text{Gamma}(3,.01) \quad  n = 1, 2, 3\\
    \sigma_{5n} \sim \text{Gamma}(3,.01) \quad n = 1, 2, 3
\end{align*}

For the initial state distribution vector, it was assumed
 $ \boldsymbol{\delta} \sim \text{Dirichlet}(\boldsymbol{1}_N)$. For the reparametrized transition probabilities, we have that the conditional priors for the baseline coefficients are
\begin{align*}
    -\alpha_{0}^{(ij)} \mid \zeta_i &\sim \text{Gumbel}(\zeta_i, 1) \quad i,j \in \{1,2,3\}, \, i\neq j,
\end{align*}
with $\zeta_i \sim \text{Gumbel}(0,1),\, i\in \{1,2,3\}$. For the extended model incorporating sound stimuli as a covariate, the conditional priors for the covariate coefficients $\alpha_{1}^{(ij)}$ are
\begin{align*}
-\alpha_{1}^{(ij)}\mid \alpha_{0}^{(ij)}, \zeta_i &\sim \text{Gumbel}(\alpha_{0}^{(ij)} + \zeta_i, 1).
\end{align*}

\clearpage

\subsection{Posterior estimates for the 3-state HMM with sound stimuli as a covariate}\label{appdx:posterior_estimates_cov}

For each of the two modes identified in the baseline 3-state HMM extended with covariates in the transition probabilities, posterior median estimates and 95\% credible intervals were computed from one million iterations after correcting for label switching, as described in Section \ref{subsec:pt_applied_hmms}. The tables below present the values for PT algorithm implementation ID 1, with 95\% credible intervals shown in parentheses. Results are organized by state-dependent parameters, transition probabilities, initial state distribution, and baseline and covariate working parameters corresponding to the reparameterization introduced in Equation \ref{eq:tpm_cov}, including transition probabilities under the presence of sound stimuli ($z_t = 1$).

\begin{table}[htp!]
\centering
\subcaption*{State-dependent parameters}
\vspace{1mm}
\scalebox{.7}{
\begin{tabular}{lcccccc}
\toprule
 & \multicolumn{2}{c}{State $n = 1$} & \multicolumn{2}{c}{State $n = 2$} & \multicolumn{2}{c}{State $n = 3$} \\
Variable & Mode $\tilde{A}$ & Mode $\tilde{B}$ & Mode $\tilde{A}$ & Mode $\tilde{B}$ & Mode $\tilde{A}$ & Mode $\tilde{B}$ \\
\midrule
$\lambda_{n}$ & 0.649 (0.575,0.728) & 0.426 (0.379,0.479) & 0.037 (0.004,0.1) & 2.82 (2.571,3.063) & 3.405 (3.216,3.601) & 2.925 (2.701,3.133) \\
$\mu_{2n}$ & 150 (142,159) & 182 (172,194) & 360 (326,398) & 401 (384,417) & 523 (510,536) & 568 (554,584) \\
$\sigma_{2n}$ & 89 (82,97) & 131 (122,143) & 226 (198,262) & 118 (107,133) & 124 (115,135) & 124 (113,136) \\
$\mu_{3n}$ & 68 (63,75) & 72 (66,77) & 96 (90,102) & 105 (100,111) & 154 (147,162) & 181 (171,191) \\
$\sigma_{3n}$ & 66 (60,74) & 67 (61,73) & 36 (31,42) & 38 (34,43) & 70 (63,77) & 68 (61,77) \\
$\mu_{4n}$ & 33 (31,36) & 30 (28,33) & 80 (69,94) & 112 (107,116) & 173 (167,179) & 216 (207,222) \\
$\sigma_{4n}$ & 24 (22,27) & 22 (20,25) & 74 (62,89) & 26 (23,29) & 60 (55,65) & 42 (37,47) \\
$\mu_{5n}$ & 210 (194,227) & 314 (291,336) & 717 (670,767) & 345 (307,399) & 414 (383,448) & 524 (473,578) \\
$\sigma_{5n}$ & 154 (138,172) & 267 (242,293) & 288 (255,329) & 227 (193,284) & 293 (264,327) & 380 (333,428) \\
\bottomrule
\end{tabular}

}
%\vspace{0.1cm}
\bigskip
\subcaption*{Transition probabilities and initial state distribution}
\vspace{1mm}
\scalebox{.7}{
\begin{tabular}{lcccccc}
\toprule
 & \multicolumn{2}{c}{State $n = 1$} & \multicolumn{2}{c}{State $n = 2$} & \multicolumn{2}{c}{State $n = 3$} \\
Variable & Mode $\tilde{A}$ & Mode $\tilde{B}$ & Mode $\tilde{A}$ & Mode $\tilde{B}$ & Mode $\tilde{A}$ & Mode $\tilde{B}$ \\
\midrule
$\gamma_{1n}$ & 0.896 (0.86,0.926) & 0.859 (0.824,0.889) & 0.028 (0.011,0.052) & 0.025 (0.012,0.047) & 0.076 (0.05,0.107) & 0.115 (0.087,0.147) \\
$\gamma_{2n}$ & 0.057 (0.02,0.113) & 0.066 (0.035,0.108) & 0.75 (0.665,0.825) & 0.921 (0.877,0.953) & 0.19 (0.123,0.27) & 0.011 (0.002,0.035) \\
$\gamma_{3n}$ & 0.102 (0.068,0.143) & 0.268 (0.208,0.327) & 0.08 (0.05,0.118) & 0.041 (0.017,0.073) & 0.817 (0.769,0.859) & 0.69 (0.624,0.75) \\
$\delta_{n}$ & 0.182 (0.08,0.325) & 0.258 (0.151,0.43) & 0.169 (0.064,0.317) & 0.22 (0.106,0.354) & 0.641 (0.476,0.786) & 0.509 (0.331,0.656) \\
\bottomrule
\end{tabular}

}
\bigskip
\subcaption*{Baseline and covariate working parameters}
\vspace{1mm}
\scalebox{1}{    \begin{tabular}{lcc}
\toprule
Variable & Mode $\tilde{A}$ & Mode $\tilde{B}$ \\
\midrule
$\alpha_{12}$ & -3.482 (-4.384,-2.831) & -3.545 (-4.315,-2.92) \\
$\alpha_{13}$ & -2.474 (-2.906,-2.096) & -2.015 (-2.315,-1.717) \\
$\alpha_{21}$ & -2.576 (-3.649,-1.825) & -2.629 (-3.298,-2.104) \\
$\alpha_{23}$ & -1.374 (-1.892,-0.917) & -4.425 (-6.245,-3.239) \\
$\alpha_{31}$ & -2.082 (-2.512,-1.7) & -0.945 (-1.269,-0.649) \\
$\alpha_{32}$ & -2.321 (-2.823,-1.892) & -2.831 (-3.749,-2.177) \\
$\beta_{12}$ & -2.423 (-4.621,-0.149) & -1.02 (-4.098,0.825) \\
$\beta_{13}$ & -1.486 (-2.847,-0.333) & -0.339 (-1.178,0.31) \\
$\beta_{21}$ & -0.697 (-3.954,1.276) & -0.555 (-2.836,1.324) \\
$\beta_{23}$ & 0.253 (-0.765,1.186) & -1.034 (-5.939,2.924) \\
$\beta_{31}$ & -0.734 (-2.343,0.412) & -0.467 (-1.314,0.312) \\
$\beta_{32}$ & 0.376 (-0.692,1.288) & -0.762 (-3.957,1.479) \\
\bottomrule
\end{tabular}

}
\bigskip
\subcaption*{Transition probabilities during the presence of sound stimuli ($z_t = 1$)}
\vspace{1mm}
\scalebox{.66}{    \begin{tabular}{lcccccc}
\toprule
 & \multicolumn{2}{c}{State $n = 1$} & \multicolumn{2}{c}{State $n = 2$} & \multicolumn{2}{c}{State $n = 3$} \\
Variable & Mode $\tilde{A}$ & Mode $\tilde{B}$ & Mode $\tilde{A}$ & Mode $\tilde{B}$ & Mode $\tilde{A}$ & Mode $\tilde{B}$ \\
\midrule
$\tilde{\gamma}_{1n}$ & 0.976 (0.934,0.993) & 0.901 (0.833,0.948) & 0.003 (0,0.026) & 0.01 (0,0.052) & 0.019 (0.005,0.054) & 0.086 (0.041,0.144) \\
$\tilde{\gamma}_{2n}$ & 0.027 (0.001,0.129) & 0.041 (0.003,0.172) & 0.725 (0.553,0.863) & 0.942 (0.781,0.995) & 0.236 (0.108,0.407) & 0.005 (0,0.093) \\
$\tilde{\gamma}_{3n}$ & 0.049 (0.011,0.129) & 0.191 (0.094,0.327) & 0.118 (0.047,0.223) & 0.021 (0.001,0.122) & 0.826 (0.711,0.912) & 0.778 (0.626,0.884) \\
\bottomrule
\end{tabular}

}
\end{table}

\subsection{7-dimensional 3-state HMM parameter estimates using maximum likelihood estimation}\label{appdx:hmm_mle}

Parameter estimates from the baseline model in \citep{deruiter_multivariate_2017} were replicated by following the same procedure, namely numerically maximizing the model’s likelihood using the \texttt{nlm} function in R. The replicated results are shown below.

\begin{table}[htp!]
\centering
\subcaption*{State-dependent parameters}
\vspace{1mm}
\scalebox{1}{
\begin{tabular}{lccc}
\toprule
Variable & State $n = 1$ & State $n = 1$ & State $n = 3$ \\
\midrule
$\mu_{2n}$ & 140 (132,148) & 334 (303,365) & 516 (503,529) \\
$\sigma_{2n}$ & 80 (73,88) & 212 (186,241) & 130 (120,140) \\
$\mu_{3n}$ & 70 (64,77) & 86 (78,95) & 151 (144,158) \\
$\sigma_{3n}$ & 68 (61,76) & 55 (47,65) & 69 (63,75) \\
$\mu_{4n}$ & 32 (30,35) & 68 (58,78) & 170 (164,176) \\
$\sigma_{4n}$ & 24 (21,26) & 65 (55,77) & 60 (56,65) \\
$\mu_{5n}$ & 187 (173,202) & 675 (627,723) & 406 (375,437) \\
$\sigma_{5n}$ & 132 (119,147) & 305 (268,347) & 287 (258,318) \\
$\kappa$ & 1.005 (0.845,1.195) & 3.023 (2.505,3.65) & 0.816 (0.656,1.015) \\
$a$ & 0.977 (0.844,1.13) & 0.501 (0.425,0.591) & 1.673 (1.459,1.918) \\
$b$ & 2.106 (1.815,2.444) & 5.432 (4.166,7.083) & 1.564 (1.37,1.786) \\
$\lambda_{n}$ & 0.67 (0.594,0.757) & 0.02 (0.004,0.091) & 3.358 (3.175,3.552) \\
\bottomrule
\end{tabular}

}
%\vspace{0.1cm}
\bigskip
\subcaption*{Initial state distribution}
\vspace{1mm}
\scalebox{1.0}{
\begin{tabular}{lccc}
\toprule
Variable & State $n = 1$ & State $n = 1$ & State $n = 3$ \\
\midrule
$\gamma_{1n}$ & 0.93 (0.876,0.96) & 0.014 (0.005,0.038) & 0.055 (0.035,0.085) \\
$\gamma_{2n}$ & 0.019 (0.006,0.059) & 0.785 (0.686,0.851) & 0.196 (0.143,0.255) \\
$\gamma_{3n}$ & 0.072 (0.049,0.103) & 0.099 (0.072,0.134) & 0.829 (0.764,0.879) \\
$\delta_{n}$ & 0.168 (0.183,0.435) & 0.171 (0.077,0.094) & 0.66 (0.489,0.723) \\
\bottomrule
\end{tabular}

}
\end{table}

\clearpage

\subsection{$\hat{R}$ and effective sample size (ESS) of estimated parameters}\label{appdx:rhat_ess}

The $\hat{R}$ and ESS were computed using one million samples for each parameter in both the baseline 3-state HMM and its extension with covariates in the transition probabilities. These values were obtained after correcting for label switching in a post-sampling process. The $\hat{R}$ values were rounded to three decimal places, while the ESS values were rounded to two decimal places. Both quantities were calculated using the functions \texttt{rhat} and \texttt{ess\_basic} from the R package \texttt{posterior} (version 1.6.0).

\subsubsection*{Baseline 3-state HMM}

\begin{table}[htp!]
\centering
\subcaption*{State-dependent parameters}
\vspace{1mm}
\scalebox{1}{
\begin{tabular}{lcccccc}
\toprule
 & \multicolumn{2}{c}{State $n = 1$} & \multicolumn{2}{c}{State $n = 2$} & \multicolumn{2}{c}{State $n = 3$} \\
Variable & $\hat{R}$ & $ESS$ & $\hat{R}$ & $ESS$ & $\hat{R}$ & $ESS$ \\
\midrule
$\lambda_{n}$ & 1.001 & 3713.77 & 1.001 & 3127.21 & 1.001 & 4038.41 \\
$\mu_{2n}$ & 1.001 & 3583 & 1.001 & 6001.93 & 1.001 & 3603.27 \\
$\sigma_{2n}$ & 1.001 & 3342.97 & 1.001 & 3530.03 & 1.000 & 748704.95 \\
$\mu_{3n}$ & 1.000 & 21901.48 & 1.001 & 5168.68 & 1.001 & 3574.69 \\
$\sigma_{3n}$ & 1.000 & 259757.41 & 1.000 & 102791.06 & 1.000 & 533488.24 \\
$\mu_{4n}$ & 1.001 & 8707.18 & 1.001 & 3763.77 & 1.001 & 3230.83 \\
$\sigma_{4n}$ & 1.001 & 8772.34 & 1.001 & 3477.5 & 1.001 & 3502.17 \\
$\mu_{5n}$ & 1.001 & 3297.53 & 1.001 & 3201.78 & 1.001 & 3747.97 \\
$\sigma_{5n}$ & 1.001 & 3279.76 & 1.001 & 5933.18 & 1.001 & 4053.81 \\
\bottomrule
\end{tabular}

}
%\vspace{0.1cm}
\bigskip
\subcaption*{Initial state distribution}
\vspace{1mm}
\scalebox{1.0}{
\begin{tabular}{lcccccc}
\toprule
 & \multicolumn{2}{c}{State $n = 1$} & \multicolumn{2}{c}{State $n = 2$} & \multicolumn{2}{c}{State $n = 3$} \\
Variable & $\hat{R}$ & $ESS$ & $\hat{R}$ & $ESS$ & $\hat{R}$ & $ESS$ \\
\midrule
$\delta_n$ & 1.000 & 14639.17 & 1.000 & 42526.81 & 1.001 & 9994.35 \\
\bottomrule
\end{tabular}

}
\bigskip
\subcaption*{Baseline working parameters}
\vspace{1mm}
\scalebox{1}{    \begin{tabular}{lcc}
\toprule
Variable & $\hat{R}$ & $ESS$ \\
\midrule
$\alpha_{12}$ & 1.000 & 452751.1 \\
$\alpha_{13}$ & 1.001 & 5499.35 \\
$\alpha_{21}$ & 1.000 & 617852.98 \\
$\alpha_{23}$ & 1.001 & 3466.2 \\
$\alpha_{31}$ & 1.001 & 3710.96 \\
$\alpha_{32}$ & 1.001 & 5335.35 \\
\bottomrule
\end{tabular}

}
\end{table}

\clearpage

\subsubsection*{3-state HMM with sound stimuli covariate}

\begin{table}[htp]
\centering
\subcaption*{State-dependent parameters}
\vspace{1mm}
\scalebox{1}{
\begin{tabular}{lcccccc}
\toprule
 & \multicolumn{2}{c}{State $n = 1$} & \multicolumn{2}{c}{State $n = 2$} & \multicolumn{2}{c}{State $n = 3$} \\
Variable & $\hat{R}$ & $ESS$ & $\hat{R}$ & $ESS$ & $\hat{R}$ & $ESS$ \\
\midrule
$\lambda_{n}$ & 1.000 & 317520.9 & 1.001 & 44298.21 & 1.000 & 387285.09 \\
$\mu_{2n}$ & 1.000 & 183615.78 & 1.000 & 205753.75 & 1.000 & 247091.23 \\
$\sigma_{2n}$ & 1.000 & 126922.04 & 1.000 & 161924.27 & 1.000 & 796138.77 \\
$\mu_{3n}$ & 1.000 & 248422.18 & 1.000 & 399463.62 & 1.000 & 219407.28 \\
$\sigma_{3n}$ & 1.000 & 262077.67 & 1.000 & 350528.63 & 1.000 & 592682.15 \\
$\mu_{4n}$ & 1.000 & 251749.29 & 1.000 & 107433.96 & 1.000 & 101629.6 \\
$\sigma_{4n}$ & 1.000 & 252209.65 & 1.000 & 109726.76 & 1.000 & 222800.43 \\
$\mu_{5n}$ & 1.000 & 102849.21 & 1.000 & 91681.94 & 1.000 & 194589.05 \\
$\sigma_{5n}$ & 1.000 & 97392.6 & 1.000 & 397324.33 & 1.000 & 239858.9 \\
\bottomrule
\end{tabular}

}
%\vspace{0.1cm}
\bigskip
\subcaption*{Initial state distribution}
\vspace{1mm}
\scalebox{1.0}{
\begin{tabular}{lcccccc}
\toprule
 & \multicolumn{2}{c}{State $n = 1$} & \multicolumn{2}{c}{State $n = 2$} & \multicolumn{2}{c}{State $n = 3$} \\
Variable & $\hat{R}$ & $ESS$ & $\hat{R}$ & $ESS$ & $\hat{R}$ & $ESS$ \\
\midrule
$\delta_n$ & 1.000 & 649477.98 & 1.000 & 620897.06 & 1.000 & 560350.02 \\
\bottomrule
\end{tabular}

}
\bigskip
\subcaption*{Baseline working parameters}
\vspace{1mm}
\scalebox{1}{    \begin{tabular}{lcc}
\toprule
Variable & $\hat{R}$ & $ESS$ \\
\midrule
$\alpha_{12}$ & 1.000 & 511634.37 \\
$\alpha_{13}$ & 1.000 & 996941.52 \\
$\alpha_{21}$ & 1.000 & 462488.08 \\
$\alpha_{23}$ & 1.000 & 100783.9 \\
$\alpha_{31}$ & 1.000 & 343611.57 \\
$\alpha_{32}$ & 1.000 & 425643.67 \\
\bottomrule
\end{tabular}
}
\bigskip
\subcaption*{Baseline working parameters}
\vspace{1mm}
\scalebox{1}{    \begin{tabular}{lcc}
\toprule
Variable & $\hat{R}$ & $ESS$ \\
\midrule
$\beta_{12}$ & 1.000 & 56804.45 \\
$\beta_{13}$ & 1.000 & 294474.78 \\
$\beta_{21}$ & 1.000 & 243709.09 \\
$\beta_{23}$ & 1.000 & 363338.05 \\
$\beta_{31}$ & 1.000 & 752708.09 \\
$\beta_{32}$ & 1.000 & 481134.05 \\
\bottomrule
\end{tabular}
}
\end{table}

\end{document}